\documentclass[a4paper,12pt]{article}
\pdfoutput=1 

\usepackage{jheppub} 

\usepackage[T1]{fontenc} 

\usepackage{graphicx} 
\usepackage{amsmath}
\usepackage{amssymb}
\usepackage{mathrsfs}
\usepackage{booktabs}
\usepackage[dvipsnames]{xcolor}
\usepackage{bbm}
\usepackage{autobreak}
\usepackage{tensor}
\usepackage{bbold}
\usepackage{multicol}
\usepackage{enumitem}

\usepackage{cleveref}
\crefname{table}{Table}{Tables}
\crefname{equation}{Eq.}{Eqs.}
\crefname{appendix}{App.}{Apps.}
\crefname{section}{Sec.}{Secs.}
\crefname{figure}{Fig.}{Figs.}
\usepackage[normalem]{ulem}
\usepackage{physics}
\usepackage{slashed}

\def\eg{\textit{e.g.}}
\def\ie{\textit{i.e.}}

\allowdisplaybreaks

\newcommand*{\scr}[1]{\mathcal{#1}}

\def\scrg{{\Y}}
\renewcommand{\op}[1]{\operatorname{#1}}
\def\pd{\partial}
\def\md{\mathrm{d}}

\def\bF{\bar{F}}
\def\bV{\bar{V}}
\def\tS{\widetilde{S}}
\def\tH{\tilde{h}}
\def\tV{\widetilde{V}}
\def\b{\text{b}}
\def\c{\text{c}}
\def\Y{\mathcal{G}'}
\def\Yt{\widetilde{\mathcal{G}}'}
\newcommand{\Lag}{{\mathcal{L}}}
\def\P{\mathcal{P}}
\def\Pd{\mathcal{P}'}
\def\W{W}
\def\U{\mathcal{U}}
\def\K{\mathcal{K}}
\def\intmdx{}

\newcommand*{\ca}[1]{{\color{Maroon} #1}}
\newcommand*{\cb}[1]{{#1}}
\newcommand*{\cc}[1]{{#1}}

\newcommand\Tstrut{\rule{0pt}{2.6ex}}         
\newcommand\Bstrut{\rule[-1.2ex]{0pt}{0pt}}   
\renewcommand\dfrac{\frac}
\newcommand\Hline{\hline\hline}

\preprint{
\vspace{-8pt}
\begin{flushright}\end{flushright}
}

\title{The Geometric Universal One-Loop Effective Action}

\author[a]{Xu-Xiang~Li,}
\author[b]{Xiaochuan~Lu,}
\author[a]{and Zhengkang~Zhang}

\affiliation[a]{Department of Physics and Astronomy, University of Utah, Salt Lake City, UT 84112, USA}
\affiliation[b]{Department of Physics, University of California, San Diego, La Jolla, CA 92093, USA}

\emailAdd{xuxiang.li@utah.edu}
\emailAdd{xil224@ucsd.edu}
\emailAdd{z.k.zhang@utah.edu}

\abstract{
We derive universal formulae for integrating out heavy degrees of freedom in scalar field theories up to one-loop level in terms of covariant quantities associated with the geometry of the field manifold. The universal matching results can be readily applied to phenomenologically interesting extensions of the Standard Model, as we demonstrate using a singlet scalar example. We also discuss the role of field redefinitions in effective field theory matching and simplifications resulting from going to a field basis where interactions are encoded in a nontrivial metric on the field manifold.
}

\begin{document}
\maketitle
\flushbottom
\newpage

\section{Introduction}
\label{sec:Introduction}

The Lagrangian formulation of quantum field theories have a huge amount of redundancies due to field redefinitions. Denoting a generic field by $\varphi$, we know that on-shell amplitudes (hence physical observables) are invariant under $\varphi\to\widetilde\varphi = f \bigl(\varphi\,,\, \partial_\mu\varphi\,,\, \partial_\mu\partial_\nu\varphi \,,\, \dots\bigr)$ with some mild restrictions on $f$ \cite{Borchers1960, Chisholm:1961tha, Kamefuchi:1961sb, tHooft:1973wag, Arzt:1993gz, Epstein2008, Manohar:2018aog, Criado:2018sdb, Criado:2024mpx}. In effective field theories (EFTs) with irrelevant operators such as the Standard Model Effective Field Theory (SMEFT) and Higgs Effective Field Theory (HEFT), field redefinitions relate different operator bases, and ambiguities associated with basis choice often complicate interpretations of (experimental or theoretical) bounds on any specific operator.

If one restricts to nonderivative field redefinitions, $\widetilde \varphi = f (\varphi)$, there is a straightforward geometric interpretation: these are coordinate transformations on the field manifold \cite{Coleman:1969sm, Callan:1969sn, Honerkamp:1971sh, Volkov:1973vd, Tataru:1975ys, Alvarez-Gaume:1981exa, Alvarez-Gaume:1981exv, Vilkovisky:1984st, DeWitt:1984sjp, Gaillard:1985uh, DeWitt:1985sg, Georgi:1991ch}. There is a natural candidate for the metric on this manifold, $g_{ij}(\varphi)$, defined from the two-derivative terms in the Lagrangian~\cite{Alonso:2015fsp, Alonso:2016oah}:
\begin{equation}
\mathcal{L} = -V(\varphi) +\frac{1}{2} \,g_{ij}(\varphi) (\partial_\mu\varphi^i)(\partial^\mu\varphi^j) + \mathcal{O}(\partial^4) \,,
\label{eq:L}
\end{equation}
where $i,j$ are flavor indices (one can also promote $\partial_\mu$ to gauge covariant derivatives). It has been shown that on-shell amplitudes can be written in manifestly covariant forms involving tensors like the Riemann curvature $R_{ijkl}$ and its covariant derivatives~\cite{Nagai:2019tgi, Cohen:2020xca, Cohen:2021ucp, Alonso:2021rac, Cheung:2021yog, Helset:2022tlf, Alonso:2023upf}. Meanwhile, precision measurements (\eg\ of the Higgs boson) are endowed with a basis-independent interpretation as measuring the curvature of the field manifold~\cite{Alonso:2015fsp, Cohen:2020xca}.\footnote{
There are also generalizations of EFT geometry that accommodate derivative field redefinitions; see Refs.~\cite{Cohen:2022uuw, Cheung:2022vnd, Cohen:2023ekv, Craig:2023wni, Craig:2023hhp, Alminawi:2023qtf, Cohen:2024bml, Lee:2024xqa} for recent developments.
}

Field space geometry has proved fruitful in obtaining many new results in recent years, including \eg\ basis-independent criteria for distinguishing SMEFT vs.\ HEFT~\cite{Alonso:2015fsp, Alonso:2016oah, Cohen:2020xca, Banta:2021dek}, all order in $\frac{v^2}{\Lambda^2}$ expressions for precision electroweak and Higgs observables~\cite{Helset:2020yio,Hays:2020scx}, and new soft theorems for general scalar theories~\cite{Cheung:2021yog, Derda:2024jvo}. It has also been used to facilitate the calculation of anomalous dimensions of SMEFT operators~\cite{Helset:2022pde, Assi:2023zid, Jenkins:2023rtg, Jenkins:2023bls}, which correspond to the {\it divergent} terms in the quantum effective action. Nevertheless, the geometric framework is yet to be applied to obtain the {\it finite} terms that arise in EFT matching (where a subset of field coordinates corresponding to heavy particles are integrated out). In this paper, we fill in this gap.

In parallel with the development of field space geometry, recent years have also seen a renaissance of functional methods for EFT matching~\cite{Henning:2014gca, Henning:2014wua, Henning:2016lyp, Fuentes-Martin:2016uol, Zhang:2016pja, Cohen:2019btp, Cohen:2020fcu, Cohen:2020qvb, Fuentes-Martin:2020udw, Fuentes-Martin:2022jrf, Fuentes-Martin:2023ljp, Born:2024mgz} based on the covariant derivative expansion (CDE) technique~\cite{Gaillard:1985uh, Chan:1986jq, Cheyette:1987qz}. The basic idea is that, instead of equating individual amplitudes in the UV theory and the EFT (\eg\ computed using Feynman diagrams), we can systematically solve the matching condition by equating the one-particle-irreducible (1PI) effective action of the EFT with the one-light-particle-irreducible (1LPI) effective action of the UV theory, which are the generating functionals of all (off-shell) amplitudes involving the light fields. In fact, when derived in this way, the one-loop EFT action has a universal structure. In particular, the loop integral part of the calculation is done once and for all, resulting in a Universal One-Loop Effective Action (UOLEA)~\cite{Henning:2014wua,Drozd:2015rsp}. Starting from a UOLEA, one-loop matching calculations for specific UV theories are reduced to elementary evaluations of matrix traces.

The UOLEA program has progressed from the initial small set of universal terms from heavy scalar loops~\cite{Drozd:2015rsp} to include mixed heavy-light scalar loops~\cite{Ellis:2016enq, Ellis:2017jns} and heavy fermion loops~\cite{Summ:2018oko, Kramer:2019fwz, Ellis:2020ivx, Angelescu:2020yzf}, and been extended to curved spacetime~\cite{Larue:2023uyv}, thereby expanding the range of UV theories for which such a streamlined matching procedure can be applied. However, further extending the UOLEA to incorporate derivative interactions is tedious with conventional methods. Field space geometry offers a natural framework to efficiently organize functional matching and derive a UOLEA for UV theories of the form \cref{eq:L} (truncated at the two-derivative level), where derivative interactions are associated with a nontrivial metric.

As a first step of showcasing the application of field space geometry for the derivation of UOLEA, we focus on scalar theories. Here, a nontrivial metric (hence derivative interactions) signals that the ``UV theory'' contains irrelevant operators. This can happen either when the UV theory is nonrenormalizable (so we are matching an EFT onto a lower-energy EFT, such as from SMEFT to Low-energy Effective Field Theory below the weak scale~\cite{Jenkins:2017jig}) or when the UV theory is renormalizable but written in a basis that contains irrelevant operators with nonzero coefficients. A simple example of the latter scenario is the linear sigma model written in spherical coordinates (\ie\ in a HEFT-like basis); we will see in Sec.~\ref{sec:sigma} how the geometric formalism helps us organize the one-loop matching calculation and results. More generally, there can be advantages to performing field redefinitions in the UV theory to trade some of the nonderivative interactions for derivative interactions; we will demonstrate this in Sec.~\ref{sec:singlet} with a phenomenologically interesting model, the singlet scalar extended Standard Model, and see that going to a field basis with derivative interactions reduces the number of UOLEA terms that need to be evaluated for one-loop matching up to dimension 6. Given that a similar field redefinition underlies the recent symmetry explanation of a magic zero~\cite{Craig:2021ksw} -- the surprising cancellation of one-loop dimension-6 dipole operator in SMEFT from integrating out heavy vector-like fermions~\cite{Arkani-Hamed:2021xlp} -- we are hopeful that the geometric formalism which naturally deals with derivative interactions can be useful for understanding more magic zeroes~\cite{DelleRose:2022ygn, Bao:2024zzc} and shedding light on hidden symmetry structures of EFTs.

The rest of this paper is organized as follows. We begin in Sec.~\ref{sec:geo/action} by reviewing the calculation of the central object in functional matching, the 1PI effective action, in the framework of field space geometry, and demonstrate the advantage of organizing the calculation in terms of geometric quantities using a simple example. Next, in Sec.~\ref{sec:geo/CDE}, we outline the matching procedure in the geometric framework. Starting from a UV Lagrangian containing up to two derivatives, the goal is to integrate out a subset of field coordinates corresponding to heavy particles to arrive at an EFT on the submanifold spanned by the light field coordinates. This procedure explicitly breaks the full covariance of the UV field manifold while preserving the covariance of the EFT submanifold. We apply this procedure to the simplest setting of $O(N)$ sigma models in Sec.~\ref{sec:sigma}, where the EFT operators are organized by the number of derivatives. We derive the UOLEA up to 4-derivative terms and obtain explicit matching result for the special case of the linear sigma model. In Sec.~\ref{sec:guolea}, we consider general UV theories with up to two derivatives where the EFT operators are organized by canonical dimension, and derive the UOLEA up to dimension 6. We then demonstrate the application of UOLEA to the matching of the singlet extended Standard Model onto SMEFT in Sec.~\ref{sec:singlet}. Finally, we summarize in Sec.~\ref{sec:summary} and discuss several future directions.

\section{Effective Action From Geometry}
\label{sec:geo/action}

We begin by reviewing the basic setup of field space geometry and the calculation of 1PI effective action in this framework~\cite{Alonso:2015fsp,Alonso:2016oah}. Consider an EFT of $N$ scalar fields $\varphi^i$ ($i=1,\dots, N$) with up to two derivatives:
\begin{equation}
{\cal L} = \frac12\, g_{ij}(\varphi) (\pd_\mu \varphi^i) (\pd^\mu \varphi^j) - V(\varphi) \,.
\label{eq:FieldSpaceGeometry/Lagrangian}
\end{equation}
Note that other two-derivative terms involving $\pd^2\varphi^i$ can be rewritten in the form shown in Eq.~\eqref{eq:FieldSpaceGeometry/Lagrangian} via integration by parts. The fields $\varphi^i$ can be viewed as a coordinate chart on the field manifold. Under a nonderivative local field redefinition $\varphi^i \to \widetilde\varphi^{\,i}(\varphi)$, $\partial_\mu \varphi^i$ transforms as a vector:
\begin{equation}
\partial_\mu \varphi^i
\quad\to\quad
\partial_\mu \widetilde\varphi^{\,i} = \left(\frac{\pd \widetilde\varphi^{\,i}}{\pd \varphi^j} \right)\partial_\mu \varphi^j \,.
\label{eq:FieldSpaceGeometry/TransformationRuleOfDerivativeTerm}
\end{equation}
As a result, $g_{ij}(\varphi)$ transforms as a symmetric $(0,2)$ tensor:
\begin{equation}
g_{ij}(\varphi)
\quad\to\quad
\widetilde g_{ij}(\widetilde\varphi) = \left(\frac{\pd \varphi^k}{\pd \widetilde\varphi^{\,i}}\right) \left(\frac{\pd \varphi^l}{\pd \widetilde\varphi^{\,j}}\right) g_{kl}\bigl(\varphi(\widetilde\varphi)\bigr) \,.
\end{equation}
In a healthy theory, we expect $g_{ij}(\varphi)$ to be positive-definite and nondegenerate in the vicinity of the vacuum, which therefore defines a metric on the field manifold. Meanwhile, the potential $V(\varphi)$ transforms as a scalar:
\begin{equation}
V(\varphi)
\quad\to\quad
\widetilde V (\widetilde\varphi) = V\bigl( \varphi(\widetilde\varphi)\bigr) \,.
\end{equation}
In this geometric framework, physical observables such as on-shell scattering amplitudes can be expressed in terms of the Riemann curvature tensor (derived from the metric $g_{ij}$), the scalar potential $V$, and their covariant derivatives, so that invariance under nonderivative field redefinitions becomes manifest~\cite{Nagai:2019tgi, Cohen:2020xca, Cohen:2021ucp, Alonso:2021rac, Cheung:2021yog, Helset:2022tlf, Alonso:2023upf}.

In the present work, our focus is on the calculation of the 1PI effective action, the central object in functional matching. In the standard background field method, we expand $\varphi^i$ around some classical background, $\varphi^i = \varphi_\b^i + \eta^i$, and obtain:
\begin{align}
S[\varphi] = S[\varphi_\b] + \eta^i\, \frac{\delta S}{\delta\varphi^i} [\varphi_\b]
+ \frac12\, \eta^i \eta^j\, \frac{\delta^2 S}{\delta\varphi^j\delta\varphi^i} [\varphi_\b]
+ \cdots \,.
\label{eq:S_standard}
\end{align}
Here and in what follows, we leave spacetime integrals implicit. Legendre transforming the logarithm of the path integral
\begin{equation}
W[J] \equiv -i \log \int \mathcal{D}\varphi \sqrt{\det g} \,e^{i(S+J_i\,\varphi^i)} \,,
\label{eqn:WJ}
\end{equation}
with respect to the source $J_i(x)$ yields the 1PI effective action:
\begin{align}
\Gamma [\varphi_\b]
\equiv W\big[ J[\varphi_\b] \big] - \varphi_\b^i J_i[\varphi_\b]
= S[\varphi_\b] +\frac{i}{2} \op{Tr}\log \left( -g^{ik}\,\frac{\delta^2 S}{\delta\varphi^j\delta\varphi^k} [\varphi_\b]\right) + \cdots \,,
\label{eq:Gamma_standard}
\end{align}
where $\varphi_\b$ is the conjugate variable of $J$ in the Legendre transform, defined as usual: $\varphi_b^i[J] \equiv \frac{\delta W}{\delta J_i}$. In the last expression of \cref{eq:Gamma_standard}, the two terms represent tree and one-loop level contributions, respectively. Note that the path integral measure is accompanied by $\sqrt{\det g}$ which results in the appearance of the inverse metric $g^{ik}$ in the one-loop level term in Eq.~\eqref{eq:Gamma_standard}. This inverse metric factor can be dropped because it only gives rise to scaleless integrals which vanish in dimensional regularization, but it turns out convenient to keep it in our calculation.

We can covariantize \cref{eq:S_standard,eq:Gamma_standard} by using geodesic coordinates $\xi^i = \frac{\md \gamma^i}{\md \lambda}(0)$ instead of $\eta^i$ to parameterize fluctuations of $\varphi^i$. Here $\gamma(\lambda)$ is the geodesic from $\varphi_\b$ to $\varphi$:
\begin{align}
\frac{\md^2 \gamma^i}{\md \lambda^2} + \Gamma^i_{jk} \frac{\md \gamma^j}{\md \lambda} \frac{\md \gamma^k}{\md \lambda} = 0 \,,
\qquad \gamma(0) = \varphi_\b \,,
\qquad \gamma(1) = \varphi \,,
\end{align}
where $\Gamma^i_{jk}$ is the Christoffel symbol of the Levi-Civita connection induced by $g_{ij}$ (not to be confused with the 1PI effective action which has the same standard notation $\Gamma$ but does not carry indices):
\begin{align}
\Gamma^i_{jk} = \frac{1}{2}\,g^{il} \bigl( g_{lj,k} +g_{lk,j} -g_{jk,l} \bigr) \,.
\end{align}
Solving the geodesic equation perturbatively, we see that:
\begin{align}
\varphi^i = \varphi_\b^i + \xi^i - \frac12\, \Gamma^i_{jk}(\varphi_\b)\, \xi^j \xi^k + O(\xi^3) \,.
\label{eq:FieldSpaceGeometry/RiemannNormalCoordinatesTransformation}
\end{align}
Therefore, in terms of $\xi^i$, the expansion of the action becomes:
\begin{align}\label{eq:FieldSpaceGeometry/ActionExpansion}
S[\varphi] = S[\varphi_\b] + \xi^i\, \nabla_i S[\varphi_\b] + \frac12\, \xi^i \xi^j\, \nabla_j \nabla_i S[\varphi_\b] + \cdots,
\end{align}
where $\nabla_i$ denotes the covariant functional derivative:
\begin{align}
\nabla_i S = \frac{\delta S}{\delta \varphi^i}\,, \qquad \nabla_j \nabla_i S = \frac{\delta^2 S}{\delta \varphi^j \delta\varphi^i} - \Gamma^k_{ji} \,\frac{\delta S}{\delta \varphi^k} \,.
\end{align}
Now consider a geometrized version of $W[J]$ in \cref{eqn:WJ}:
\begin{equation}
W_\text{geo}[J] \equiv -i \log \int \mathcal{D}\varphi \sqrt{\det g} \,e^{i(S+J_i (\varphi_\b^i + \xi^i))} \,.
\label{eqn:WgeoJ}
\end{equation}
The difference is that in $W_\text{geo}[J]$ the source variable $J_i$ is coupled to $\varphi_\b^i + \xi^i$ instead of $\varphi^i$ (see \cref{eq:FieldSpaceGeometry/RiemannNormalCoordinatesTransformation}). Legendre transforming  $W_\text{geo}[J]$ then gives a geometrized 1PI effective action:
\begin{align}
\Gamma_\text{geo} [\varphi_\b] = S[\varphi_\b] +\frac{i}{2} \op{Tr}\log \bigl(-g^{ik}\,\nabla_j\nabla_k S [\varphi_\b]\bigr) + \cdots \,.
\label{eq:Gamma_covariant}
\end{align}
Generally, $\Gamma_\text{geo}[\varphi_\b]$ in Eq.~\eqref{eq:Gamma_covariant} and $\Gamma[\varphi_b]$ in Eq.~\eqref{eq:Gamma_standard} are different generating functionals for the off-shell amplitudes, but they yield the same on-shell amplitudes \cite{Alonso:2016oah, Manohar:2018aog, Cohen:2023ekv}.

For the Lagrangian in Eq.~\eqref{eq:FieldSpaceGeometry/Lagrangian}, the covariant functional derivatives of the action read:\footnote{As is conventional, we leave the $\delta^d(x-y)$ factor on the right-hand side of Eq.~\eqref{eq:nabla2_S} implicit (the two functional derivatives are taken with respect to $\varphi^j(x)$ and $\varphi^k(y)$, respectively).}
\begin{subequations}
\label{eq:nabla_S}
\begin{align}
\nabla_i S = & - g_{ij} \bigl({\scr D}^\mu (\pd_\mu \varphi)\bigr)^j - V_{;i}\,, \\[5pt]
g^{ik} \,\nabla_j \nabla_k S = & -({\scr D}^2 )\indices{^i_j} - R\indices{^i_{kjl}}(\pd_\mu \varphi^k) (\pd^\mu \varphi^l) - V\indices{_;^i_{\;\,j}} \,,\label{eq:nabla2_S}
\end{align}
\end{subequations}
where
\begin{align}
R\indices{^i_{kjl}} = \Gamma^i_{kl,j} - \Gamma^i_{kj,l} + \Gamma^i_{jm} \Gamma^m_{kl} - \Gamma^i_{lm} \Gamma^m_{kj}
\end{align}
is the Riemann curvature tensor, and the covariant derivative ${\scr D}_\mu$ acts on a general vector $v$ in the following way:
\begin{align}
{(\scr D}_\mu v)^i = \pd_\mu v^i + (\pd_\mu \varphi)^j \Gamma^i_{jk} v^k \,.
\label{eq:D}
\end{align}
In other words, ${\scr D}_\mu$ is a covariant derivative on the spacetime manifold pulled back from the field manifold by the field $\varphi$. With \cref{eq:nabla_S}, the geometrized 1PI effective action $\Gamma_\text{geo}[\varphi_\b]$ (up to one-loop level) takes a manifestly covariant form.\footnote{One can also extend the covariant expression of the 1PI effective action to higher loop orders~\cite{Alonso:2022ffe}.}
One can then, for example, extract the UV divergences from $\Gamma_\text{geo}[\varphi_\b]$ and compute RG running of EFT operators in terms of covariant derivatives of the potential and Riemann curvature~\cite{Alonso:2015fsp, Alonso:2016oah, Helset:2022pde, Assi:2023zid}. In this work, our goal is to extract additional information from $\Gamma_\text{geo}[\varphi_\b]$ by computing also the finite terms in the hard region expansion when $V$ involves disparate mass scales. In other words, we compute the EFT matching coefficients upon integrating out a subset of fields from \cref{eq:FieldSpaceGeometry/Lagrangian}.

\subsection*{Single-Flavor Example}

Before getting into the details of EFT matching, let us briefly illustrate how the covariantization above simplifies the evaluation of the 1PI effective action. We consider a simple example --- a $\mathbb{Z}_2$-symmetric single-flavor theory:
\begin{equation}
\Lag = \frac12\, g_{\varphi\varphi}(\varphi) ( \partial_\mu \varphi ) ( \partial^\mu \varphi ) - V(\varphi) \,,
\label{eqn:LagSingleScalar}
\end{equation}
where $g_{\varphi\varphi}(\varphi) = g_{\varphi\varphi}(-\varphi)$, $V(\varphi) = V(-\varphi)$. Separating out the mass term from the potential, $V(\varphi) = \frac{1}{2}\,M^2\,\varphi^2 + \cdots$, we can write the one-loop contribution to the (original) 1PI effective action \cref{eq:Gamma_standard} as:
\begin{equation}
\Gamma^{[1]} = \frac{i}{2} \op{Tr} \log \biggl( -g^{\varphi\varphi} \, \frac{\delta^2 S}{\delta\varphi^2} \biggr)
= \frac{i}{2} \op{Tr} \log \left( P^2 - Z_\mu P^\mu - M^2 - U \right) \,,
\label{eqn:Gamma1}
\end{equation}
where
\begin{subequations}\label{eqn:Components}
\begin{align}
P_\mu &= i\partial_\mu \,, \qquad Z_\mu = -2i\, \Gamma_{\varphi\varphi}^\varphi (\partial_\mu \varphi) \,, \\[8pt]
U &= \U + \frac12 \left( P_\mu Z^\mu \right) - \frac14\, Z_\mu Z^\mu + E_\varphi\, g^{\varphi\varphi} \,\Gamma^\varphi_{\varphi\varphi} \,, \label{eqn:UinXJ} \\[5pt]
\U &= g^{\varphi\varphi} V_{;\varphi\varphi} - M^2 \,, \qquad
E_\varphi = -\frac{\delta S}{\delta\varphi} = g_{\varphi\varphi} \bigl(\scr{D}^\mu(\partial_\mu\varphi)\bigr) +V_{,\varphi} \,,
\end{align}
\end{subequations}
and we have dropped an irrelevant constant. Here and in what follows, a derivative operator $P_\mu = i\partial_\mu$ acts on everything to its right (and in that sense we call it ``open'') unless when it is enclosed by parentheses which can be equivalently written as a commutator, \eg\ $(P_\mu Z^\mu) \equiv [P_\mu, Z^\mu] = i(\partial_\mu Z^\mu)$. Note that the $\mathbb{Z}_2$ symmetry implies that $\U$ starts at mass dimension 2, and $Z_\mu$ starts at mass dimension 3. In what follows, we use square brackets to denote the minimum operator dimension, \eg\ $[\varphi]=1$. The minimum operator dimensions of various quantities in our simple example are summarized in \cref{tab:MinOpdimSimple}.

\begin{table}[t]
\renewcommand{\arraystretch}{1.0}
\setlength{\arrayrulewidth}{.2mm}
\setlength{\tabcolsep}{1em}
\centering
\begin{tabular}{cccccc}
\toprule
$[P_\mu]$ & $[\Gamma^\varphi_{\varphi\varphi}]$ & $[Z_\mu]$ & $[U]$ & $[E_\varphi]$ & $[\U]$ \\[2pt]
\midrule
1 & 1 & 3 & 2 & 1 & 2 \\
\bottomrule
\end{tabular}
\caption{\label{tab:MinOpdimSimple}
Minimum operation dimensions of various quantities in the $\mathbb{Z}_2$-symmetric single-flavor theory.}
\end{table}

We can evaluate the functional trace in \cref{eqn:Gamma1} as a power series in $\frac{1}{M^2}$ using CDE methods. Note that in addition to $P^2 - M^2 - U$, this expression contains another term $Z_\mu P^\mu$ with an open covariant derivative, which complicates the CDE evaluation. From the general results in Ref.~\cite{Cohen:2019btp} and the minimum operator dimensions listed in \cref{tab:MinOpdimSimple}, we obtain\footnote{We use dimensional regularization and $\overline{\text{MS}}$ scheme throughout this work.}
\begin{align}
\frac{i}{2}\log\det\left( P^2 - Z_\mu P^\mu - M^2 - U \right) &= \int \dd^4x\, \frac{1}{16\pi^2}\, \tr \Bigg\{
M^2\, \frac12\, (1-L) \left( U + \frac14\, Z_\mu Z^\mu \right)
\notag\\[5pt]
&\hspace{-125pt}
- \frac14\, L \bigg[ U^2 - U \left( P_\mu Z^\mu \right) \bigg]
- \frac{1}{M^2}\, \frac{1}{12} \bigg[ U^3 + \frac12 \left( P_\mu U \right)^2 \bigg]
\Bigg\}  + (\text{dim-8}) \,,
\label{eqn:CDEP2JPM2U}
\end{align}
where $L\equiv\log(M^2/\mu^2)$, and we have used $[P_\mu, P_\nu]=0$. Now substituting in \cref{eqn:UinXJ}, we obtain
\begin{align}
\Gamma^{[1]} &= \int \dd^4x\, \frac{1}{16\pi^2}\, \tr \Bigg\{
M^2\, \frac12\, (1-L)\; \U
- \frac14\, L\; \U^2
- \frac{1}{M^2}\, \frac{1}{12} \bigg[\, \U^3 + \frac12 \left( P_\mu \,\U \right)^2 \bigg]
\Bigg\} 
\notag\\[5pt]
&\quad
+ (\text{terms involving $E_\varphi$}) + (\text{dim-8}) \,.
\label{eqn:GammaSimple}
\end{align}
The terms involving the equation of motion (EOM) $E_\varphi$ can be eliminated by a field redefinition under which on-shell physics is unchanged. We see that the end result only depends on the geometric scalar function $\U$, and terms involving the non-covariant quantity $Z_\mu= -2i\, \Gamma_{\varphi\varphi}^\varphi (\partial_\mu \varphi)$ have canceled out.

We can in fact arrive at the same result more easily by computing the geometric version of the 1PI effective action in a manifestly covariant way. The one-loop level term in \cref{eq:Gamma_covariant} is
\begin{equation}
\Gamma^{[1]}_\text{geo} = \frac{i}{2} \op{Tr}\log \bigl(-g^{\varphi\varphi}\,\nabla_\varphi^2 S \bigr) = \frac{i}{2} \op{Tr} \log \left( \scr{P}^2 - M^2 - \U \right) \,,
\label{eqn:Gamma1Geo}
\end{equation}
where
\begin{equation}
{\scr P}_\mu \equiv P_\mu - \frac12\, Z_\mu \,.
\label{eqn:PGeo}
\end{equation}
By absorbing $Z_\mu$ into the new (covariant) derivative operator $\P_\mu$, we now have a functional trace that is simpler to evaluate with CDE. From the general results in Ref.~\cite{Henning:2014wua} and the minimum operator dimensions listed in \cref{tab:MinOpdimSimple}, we obtain
\begin{align}
\frac{i}{2}\log\det\left( \P^2 - M^2 - \U \right) &= \int \dd^4x\, \frac{1}{16\pi^2}\, \tr \Bigg\{
M^2\, \frac12\, (1-L)\; \U
- \frac14\, L\; \U^2
\notag\\[5pt]
&\hspace{40pt}
- \frac{1}{M^2}\, \frac{1}{12} \bigg[\, \U^3 + \frac12 \left( \P_\mu\, \U \right)^2
\bigg] \Bigg\} + (\text{dim-8}) \,,
\label{eqn:CDEP2M2X}
\end{align}
where we have used $[\P_\mu, \P_\nu]=0$. Noting that $(\P_\mu \,\U) = (P_\mu \,\U) -\frac{1}{2} [Z_\mu, \U] = (P_\mu \,\U)$, we see that Eq.~\eqref{eqn:CDEP2M2X} directly leads to the terms written out in \cref{eqn:GammaSimple}. The difference between the original 1PI effective action $\Gamma$ and the covariant version $\Gamma_\text{geo}$ is proportional to the EOM $E_\varphi$, so $\Gamma$ and $\Gamma_\text{geo}$ contain the same on-shell physics~\cite{Alonso:2016oah}.

\section{Geometrizing Functional Matching}
\label{sec:geo/CDE}

In this section, we describe the general procedure and the techniques involved for performing EFT matching calculations in a geometrically covariant manner. Consider a theory $\mathcal{L}[\varphi]$ with a mass hierarchy:
\begin{align}
\varphi^i = (\Phi^A, \phi^a) \,,
\qquad \text{with} \quad
M_\Phi \gg m_\phi \,.
\end{align}
We use capital (lowercase) indices starting from the beginning of the alphabet, $A, B, \dots$ ($a,b,\dots$), for the heavy (light) fields $\Phi$ ($\phi$), and use lowercase indices starting from the middle of the alphabet, $i,j, \dots$, for generic fields $\varphi$ which can be either heavy or light. We would like to integrate out $\Phi$ to obtain an EFT $\mathcal{L}_\text{EFT}[\phi]$, whose power counting is set by $1/M_\Phi$.

A familiar approach to deriving $\mathcal{L}_\text{EFT}[\phi]$ is to match a collection of amplitudes among the light fields $\phi$ between the full theory and the EFT;
see \eg\ Ref.~\cite{Carmona:2021xtq}. An equivalent but more efficient approach is to equate the generating functionals \cite{Georgi:1991ch, Georgi:1993mps}, the 1PI effective action of the EFT $\Gamma_\text{EFT}[\phi_\b]$ and the 1LPI effective action of the full theory $\Gamma_\text{L}[\phi_\b]$ (obtained by setting $\Phi_\b$ in the 1PI effective action of the full theory $\Gamma[\Phi_\b, \phi_\b]$ to its EOM solution $\Phi_\c[\phi_b]$, \ie\ $\Gamma_\text{L}[\phi_\b] = \Gamma\bigl[\Phi_\c[\phi_\b],\phi_\b\bigr]$). In full generality, this latter approach yields
\begin{align}
\mathcal{L}_\text{EFT} [\phi] = \mathcal{L}_\text{EFT}^{[0]} [\phi] + \mathcal{L}_\text{EFT}^{[1]} [\phi] + \cdots \,,
\end{align}
where tree and one-loop level results (labeled by superscripts ``$[0]$'' and ``$[1]$,'' respectively) are given by~\cite{Fuentes-Martin:2016uol, Zhang:2016pja, Cohen:2020fcu}
\begin{subequations}
\begin{align}
\int \md^d x \,\mathcal{L}_\text{EFT}^{[0]} [\phi] &=
\Gamma^{[0]} \bigl[\Phi_\c[\phi],\phi\bigr] \,,\\[5pt]
\int \md^d x \,\mathcal{L}_\text{EFT}^{[1]} [\phi] &=
\Gamma^{[1]} \bigl[\Phi_\c[\phi],\phi\bigr]\Bigr|_\text{hard} \,,
\end{align}
\end{subequations}
where $d=4-\epsilon$ in dimensional regularization. Here ``hard'' means expanding the loop integrand in the hard region \cite{Beneke:1997zp, Smirnov:2002pj}, where the loop momentum $q\sim\mathcal{O}(M)$, before performing the integration. Recent works have utilized CDE methods~\cite{Gaillard:1985uh, Chan:1986jq, Cheyette:1987qz, Henning:2014wua} to compute matching results in a gauge-covariant way. The same CDE method will allow us to derive EFT operators from a UV theory of the form \cref{eq:FieldSpaceGeometry/Lagrangian} in a geometrically covariant way (where we have turned off gauge interactions for simplicity), starting from the covariant expression for the 1PI effective action in \cref{eq:Gamma_covariant}.

To set up the matching calculation, suppose
\begin{align}
V(\varphi) = \frac{1}{2}\,M^2\,\delta_{AB} \,\Phi^A\Phi^B + \W(\varphi) \,.
\label{eq:V_sep}
\end{align}
Separating out the mass terms of heavy fields explicitly breaks the covariance on the full field manifold. However, covariance on the EFT submanifold remains: $\W(\varphi)$ in \cref{eq:V_sep} is a scalar as far as field redefinitions within the EFT submanifold are concerned:
\begin{align}
\W(\Phi, \phi) \;\to\; \widetilde \W(\Phi, \widetilde\phi) = \W\bigl( \Phi, \phi(\widetilde\phi)\bigr) \,.
\label{eqn:phiRedef}
\end{align}
Note that in \cref{eq:V_sep} we have assumed any mass splitting among the heavy fields must be much less than the overall heavy mass scale $M$.

\subsection*{Tree-Level Matching}

Matching at tree level amounts to solving the EOM of the heavy fields:
\begin{multline}
\nabla_A S \bigl[\Phi_\c[\phi],\phi\bigr] =  \Bigl[- g_{Aj} \bigl({\scr D}^\mu (\pd_\mu \varphi)\bigr)^j - V_{;A}\Bigr]\Bigr|_{\Phi = \Phi_\c[\phi]} = 0 \\[8pt]
\Longrightarrow \quad \Phi_\c^A = -\frac{1}{M^2}\,\delta^{AB} \,\Bigl[ \W_{,B} +g_{Bj}\bigl({\scr D}^\mu (\pd_\mu \varphi)\bigr)^j \Bigr] \Bigr|_{\Phi = \Phi_\c} \,.
\label{eq:eom} 
\end{multline}
In practice, \cref{eq:eom} is solved iteratively as an expansion in powers of the light fields $\phi$ and derivatives. Therefore, the solution $\Phi_\c[\phi]$ is an infinite series of local operators of the light fields $\phi$, which we can substitute into the UV Lagrangian to obtain the tree-level EFT Lagrangian (truncated at any finite operator dimension):
\begin{align}
\mathcal{L}_\text{EFT}^{[0]} [\phi] = \mathcal{L} \bigl[\Phi_\c[\phi],\phi\bigr]\,.
\label{eq:L0EFT}
\end{align}
Note that \cref{eq:eom} and consequently \cref{eq:L0EFT} are manifestly invariant on the EFT submanifold; $\Phi_\c^A$ carries a heavy flavor index but is a scalar as far as field redefinitions within the EFT submanifold are concerned (similar to \cref{eqn:phiRedef}).

\subsection*{One-Loop Matching}

Matching at one-loop level requires evaluating the functional trace in \cref{eq:Gamma_covariant}. From \cref{eq:nabla2_S,eq:V_sep} we find
\begin{align}
g^{ik} \,\nabla_j \nabla_k S = ({\scr P}^2 -{\bf M}^2 -\U)\indices{^i_j} \,,
\end{align}
where
\begin{subequations}
\begin{align}
({\scr P}_\mu)^i{}_j &\equiv
i\, ({\scr D}_\mu)^i{}_j = i\, \delta^i_j\, \pd_\mu + i\, (\pd_\mu \varphi^k)\, \Gamma^i_{kj} \,, \label{eqn:scrPdef} \\[8pt]
\hspace{-20pt}
\bigl({\bf M}^2\bigr)\indices{^i_j} &\equiv
\delta^i_A \delta_j^A \, M^2\,, \qquad \text{\ie} \quad {\bf M}^2 = \text{diag}(M^2, \dots, M^2, 0, \dots, 0) \,, \\[5pt]
\U\indices{^i_j} &\equiv
R\indices{^i_{kjl}} \bigl(\partial_\mu \varphi^k\bigr) \bigl(\partial^\mu \varphi^l\bigr) + \W\indices{_;^i_{\;j}} + \Bigl[\bigl(g^{iA}-\delta^{iA}\bigr) \,\delta_{Aj} -g^{ik}\, \Gamma^A_{jk} \,\delta_{AB} \,\Phi^B \Bigr] M^2 \,.
\end{align}
\end{subequations}
Here and in what follows, we use semicolon to represent the action of the same linear operator as the covariant derivative, even for nontensorial objects, \eg\ $\W\indices{_;^i_{\;j}}$ is defined as $g^{ik} ( \W_{,kj} -\Gamma^l_{kj} \W_{,l})$. The one-loop EFT action is therefore given by
\begin{align}
\int \md^d x \,\mathcal{L}_\text{EFT}^{[1]} [\phi] = \frac{i}{2} \op{Tr}\log \bigl( {\scr P}^2 -{\bf M}^2 -\U \bigr)\Bigr|_{\Phi = \Phi_\c[\phi], \,\text{hard}} \,.
\label{eq:LEFT1}
\end{align}
This is analogous to the expression in standard functional matching calculations for UV theories without derivative interactions. The new features here are that ${\scr P}_\mu$ is the geometric covariant derivative (as opposed to partial or gauge covariant derivative, usually denoted by $P_\mu$) and, as in the tree-level calculation, covariance on the EFT submanifold is preserved.

\subsection*{Evaluation With CDE}

When evaluating a functional trace with operators appearing in the logarithm, it is typically easier to differentiate with respect to some parameter, \eg\ $M^2$, to turn the logarithm into an inverse. However, this is straightforward only when the argument of the logarithm is (block-)diagonal. In the present case of \cref{eq:LEFT1}, we generically have\footnote{When the expressions are written in matrix form, it is assumed that each object has an upper index followed by a lower index, \eg\ the block $\U_{HH}$ has entries $\U^A{}_B$, and similarly for the other matrices introduced below (see \cref{eq:Pd_K_def}).}
\begin{align}\label{eq:geo/u}
\U = 
\begin{pmatrix}
\,\U_{HH} & \U_{HL} \\
\,\U_{LH} & \U_{LL} 
\end{pmatrix} \,,
\end{align}
with nonzero off-diagonal elements, $\U_{HL} \,,\, \U_{LH}\ne 0$, between heavy ($H$) and light ($L$) blocks. The same is also true for the geometric covariant derivative ${\scr P}_\mu$. To proceed, we split ${\scr P}_\mu$ into block-diagonal and the off-block-diagonal parts:
\begin{align}
{\scr P}_\mu = \Pd_\mu + \K_\mu \,,
\end{align}
with
\begin{align}
\Pd_\mu = 
\begin{pmatrix}
\bigl(\Pd_\mu\bigr)_{HH} & 0 \\
0 & \bigl(\Pd_\mu\bigr)_{LL}
\end{pmatrix} \,,
\qquad
\K_\mu = 
\begin{pmatrix}
0 & \bigl(\K_\mu\bigr)_{HL} \\
\bigl(\K_\mu\bigr)_{LH} & 0
\end{pmatrix} \,.
\end{align}
From the definition in \cref{eqn:scrPdef}, we can read off the entries in each block:
\begin{subequations}\label{eq:Pd_K_def}
\begin{alignat}{2}
\bigl(\Pd_\mu\bigr)_{HH} \;:&&\qquad
\bigl(\Pd_\mu\bigr)\indices{^A_B} &= i\,\delta^A_B \,\partial_\mu +i\, (\partial_\mu\varphi^k)\, \Gamma^A_{kB} \,, \\[8pt]
\bigl(\Pd_\mu\bigr)_{LL} \;:&&\qquad
\bigl(\Pd_\mu\bigr)\indices{^a_b} &= i\,\delta^a_b \,\partial_\mu +i\, (\partial_\mu\varphi^k)\, \Gamma^a_{kb} \,, \\[8pt]
\bigl(\K_\mu\bigr)_{HL} \;:&&\qquad
\bigl(\K_\mu\bigr)\indices{^A_b} &= i\, (\partial_\mu\varphi^k)\, \Gamma^A_{kb} \,, \\[8pt]
\bigl(\K_\mu\bigr)_{LH} \;:&&\qquad
\bigl(\K_\mu\bigr)\indices{^a_B} &= i\, (\partial_\mu\varphi^k)\, \Gamma^a_{kB} \,.
\end{alignat}
\end{subequations}

Following Ref.~\cite{Cohen:2020fcu}, we can separate the logarithm in Eq.~\eqref{eq:LEFT1} into ``log-type'' and ``power-type'' traces, with the former containing only block-diagonal matrices:\footnote{Note that the commutator terms from applying the Baker-Campbell-Hausdorff formula, $\log (AB) = \log A + \log B + \frac{1}{2}\,[\log A, \log B] + \cdots$, vanish upon taking the trace.}
\begin{align}
&\op{Tr}\log \bigl( {\scr P}^2 - {\bf M}^2 - \U \bigr) \nonumber\\[5pt]
=\; & \op{Tr}\log \bigl( \Pd^2 - {\bf M}^2 \bigr) + \op{Tr}\log \Bigl[ 1 - \bigl(\Pd^2 - {\bf M}^2\bigr)^{-1}\, \bigl( \,\U - \K_\mu \K^\mu -\,\{\Pd_\mu, \K^\mu\}\bigr) \Bigr] \nonumber \\[5pt]
=\; & \op{Tr}\log \bigl(\Pd^2 - {\bf M}^2 \bigr) - \sum_{n=1}^\infty \frac{1}{n} \op{Tr}\Bigl[ \bigl(\Pd^2 - {\bf M}^2\bigr)^{-1} \,\bigl( \,\U - \K_\mu \K^\mu -\,\{\Pd_\mu, \K^\mu\}\bigr) \Bigr]^n \,.
\end{align}
The log-type trace can then be rewritten as the integral of a power-type trace:
\begin{align}
\op{Tr}\log \bigl(\Pd^2 - {\bf M}^2 \bigr) = 
-\int \dd M^2\, \op{Tr}_H \Big[ \bigl(\Pd^2 - {\bf M}^2 \bigr)^{-1} \Big] \,,
\end{align}
where $\op{Tr}_H$ denotes that the trace is taken over the heavy subspace.

To evaluate a power-type trace --- denote it by $\op{Tr} f \bigl(\Pd_\mu, \,\U, \,\K_\mu \bigr)$ --- we first evaluate the functional part, which amounts to introducing spacetime and momentum integrals and shifting $\partial_\mu \to \partial_\mu +iq_\mu$ with $q$ the loop momentum:
\begin{align}
\op{Tr} f\bigl(\Pd_\mu,\, \U,\, \K_\mu\bigr) = \int\md^dx \int\frac{\md^dq}{(2\pi)^d}\, \op{tr} f\bigl(\Pd_\mu-q_\mu,\, \U,\, \K_\mu\bigr) \,.
\end{align}
Next, we insert factors of $e^{\pm \Pd\cdot \partial_q}$ as in the ``original CDE'' method~\cite{Gaillard:1985uh,Cheyette:1987qz,Henning:2014wua,Cohen:2019btp}:
\begin{align}
\op{Tr} f\bigl(\Pd_\mu, \, \U,\, \K_\mu\bigr)
= \int\md^dx \int\frac{\md^dq}{(2\pi)^d}\, \op{tr} e^{ \Pd\cdot \partial_q} \,f\bigl(\Pd_\mu-q_\mu, \, \U,\, \K_\mu\bigr) \,e^{- \Pd\cdot \partial_q} \,,
\end{align}
which is legitimate since all but the first terms in $e^{\pm \Pd\cdot \partial_q} = 1\pm \Pd\cdot \partial_q + \cdots$ yield either $\partial_q \mathbb{1} = 0$ or total derivatives. The $e^{\pm \Pd\cdot \partial_q}$ insertions allow us to assemble all covariant derivatives into commutators:\footnote{Recall that taking derivative on a function amounts to evaluating a commutator: $\bigl(\partial_\mu f(x)\bigr) = \partial_\mu f(x) - f(x)\, \partial_\mu = \bigl[\partial_\mu, f(x) \bigr]$, where $\partial_\mu$ is an operator acting on everything to its right. In this notation, all derivatives must be in commutators in the final result for EFT operators. The $e^{\pm \Pd\cdot \partial_q}$ insertions here ensures this is the case in all intermediate steps.}
\begin{subequations}
\begin{align}
e^{\Pd\cdot\pd_q}\, (\Pd_\mu - q_\mu)\, e^{-\Pd\cdot\pd_q} &= -q_\mu + \sum_{n = 0}^{\infty} \frac{n+1}{(n+2)!} \bigl[\Pd_{\alpha_1}, \bigl[ \cdots [\Pd_{\alpha_n}, \Y_{\mu\nu}]\cdots\bigr]\bigr]\, \pd_q^\nu \,\pd_q^{\alpha_1} \cdots \pd_q^{\alpha_n} \nonumber \\[5pt]
&\equiv\, -q_\mu + \Yt_{\mu\nu} \,\pd_q^\nu \,, \\[10pt]
e^{\Pd\cdot\pd_q}\, \U\, e^{-\Pd\cdot\pd_q} &= \sum_{n = 0}^\infty \frac{1}{n!} \,\bigl[\Pd_{\alpha_1}, \bigl[ \cdots [\Pd_{\alpha_n}, \U]\cdots\bigr]\bigr] \,\pd_q^{\alpha_1} \cdots \pd_q^{\alpha_n} \;\equiv \,\widetilde{\U}\,, \\[10pt]
e^{\Pd\cdot\pd_q}\, \K_\mu\, e^{-\Pd\cdot\pd_q} &= \sum_{n = 0}^\infty \frac{1}{n!} \,\bigl[\Pd_{\alpha_1}, \bigl[ \cdots [\Pd_{\alpha_n}, \K_\mu]\cdots\bigr]\bigr] \,\pd_q^{\alpha_1} \cdots \pd_q^{\alpha_n} \;\equiv\, \widetilde{\K}_\mu\,, 
\end{align}
\label{eq:YUKtilde}
\end{subequations}
where $\Y_{\mu\nu} \equiv -\,[\Pd_\mu , \Pd_\nu ]$. Therefore:
\begin{align}
\op{Tr} f\bigl(\Pd_\mu, \, \U,\, \K_\mu\bigr)
= \int\md^dx \int\frac{\md^dq}{(2\pi)^d}\, \op{tr} f\bigl(-q_\mu + \Yt_{\mu\nu} \,\pd_q^\nu, \, \widetilde \U,\, \widetilde \K_\mu\bigr) \,.
\end{align}

Finally, to obtain the EFT action from Eq.~\eqref{eq:LEFT1}, we need to expand the loop integrand in the hard region $q\sim M_\Phi \gg m_\phi$. Since we have defined ${\bf M}^2$ to have zero entries in the light block, all light particle masses are included the $\U$ matrix. So the hard region expansion amounts to writing the integrand as a series in $\frac{1}{q^2-{\bf M}^2}$ (treating all other quantities as much smaller than $q^2\sim M^2$), and we obtain:\footnote{Note that the first term coming from the log-type trace is automatically over the heavy subspace, since for light fields the integral is scaleless and vanishes in dimensional regularization.}
\begin{align}
{\cal L}_{\text{EFT}}^{[1]} 
&= -\frac{i}{2} \int \dd M^2 \int \frac{\dd^dq}{(2\pi)^d}\, \op{tr} \Bigg[
\sum_{m=0}^\infty \tfrac{1}{q^2-{\bf M}^2}
\Bigl\{ \Big[ \{q^\mu , \Yt_{\mu\nu} \partial_q^\nu \} 
- ( \Yt_{\mu\nu} \partial_q^\nu )^2 \Big] \tfrac{1}{q^2-{\bf M}^2} \Bigr\}^m
\Bigg] \Biggr|_{\Phi = \Phi_\c[\phi]}
\notag\\[10pt]
&\quad
-\frac{i}{2}\, \sum_{n=1}^\infty \frac{1}{n} \int \frac{\md^dq}{(2\pi)^d}\, \op{tr} \Bigg[
\sum_{m=0}^\infty \tfrac{1}{q^2-{\bf M}^2}
\Bigl\{ \Big[ \{ q^\mu , \Yt_{\mu\nu} \partial_q^\nu \}
- ( \Yt_{\mu\nu} \partial_q^\nu )^2 \Big] \tfrac{1}{q^2 - {\bf M}^2} \Bigr\}^m
\notag\\[5pt]
&\hspace{110pt}
\times \left( \widetilde \U - \widetilde \K_\mu \widetilde \K^\mu 
+ \acomm{q_\mu - \Yt_{\mu\nu} \partial_q^\nu}{\widetilde \K^\mu} \right)
\Bigg]^n \Biggr|_{\Phi = \Phi_\c[\phi]} \,.
\label{eq:geo/eft_lag_expand}
\end{align}
Once the power counting is specified, we can enumerate terms in Eq.~\eqref{eq:geo/eft_lag_expand} up to any given order in the EFT expansion and evaluate the momentum integrals. In the next two sections, we will derive universal one-loop matching formulas (\ie\ UOLEAs) from Eq.~\eqref{eq:geo/eft_lag_expand} for $O(N)$ sigma models and for general two-derivative theories, respectively. In the former case, the EFT is organized as a derivative expansion, and we obtain terms containing up to 6 derivatives at tree level and 4 derivatives at one-loop level; in the latter case, the power counting is set by the canonical dimension, and we obtain UOLEA terms sufficient for EFT matching up to dimension 6.

\section{UOLEA for $O(N)$ Sigma Models}
\label{sec:sigma}

As a first application of the geometrized matching formalism, let us consider UV Lagrangians of the form of $O(N)$ sigma models:
\begin{equation}
{\cal L}[h, \pi] = \frac{1}{2}\, (\partial_\mu h)^2 + \frac{1}{2}\, F^2(h)\, \hat{g}_{ab}(\pi) (\partial_\mu \pi^a) (\partial^\mu \pi^b) - V(h) \,,
\label{eq:sigma/Lagrangian}
\end{equation}
where $a,b = 1, \dots, N-1$, $\hat g_{ab}(\pi) = 1+ \mathcal{O}(\pi^2)$, and
\begin{subequations}
\begin{align}
F(h) &= 1 + \bF_{,h}\, h + \frac{1}{2}\, \bF_{,hh}\, h^2 + \cdots \,, \\[5pt]
V(h) &= \frac{1}{2}\, M^2\, h^2 + \frac{1}{3!}\, \bV_{,hhh}\, h^3 + \cdots \,.
\end{align}
\end{subequations}
We use comma to denote derivatives and bar to denote evaluation at $h=0$, \eg\ $\bF_{,h}\equiv \frac{dF}{dh}\bigr|_{h=0}$. The field manifold is charted by $\{h, \pi^a\}$, with the metric given by
\begin{align}
g_{ij} = \left(
\begin{array}{cc}
1 & 0 \\
0 & F^2(h)\, \hat{g}_{ab}(\pi)
\end{array}\right).
\end{align}
The nonvanishing components of the connection are
\begin{subequations}\label{eq:sigma_connection}
\begin{align}
\Gamma^a_{bc} &= \frac{1}{2}\,\hat{g}^{ad}(\hat{g}_{dc,b} + \hat{g}_{db,c} - \hat{g}_{bc,d}) \equiv \hat{\Gamma}^a_{bc} \,,\label{eq:sigma_connection_abc} \\[6pt]
\Gamma^a_{bh} &= \frac{F_{,h}}{F}\,\delta^a_b \,, \\[8pt]
\Gamma^h_{bc} &= -FF_{,h} \,\hat{g}_{bc} \,.
\end{align}
\end{subequations}
We will integrate out the radial mode $h$ and derive the EFT Lagrangian for the Goldstone modes $\pi^a$. An important feature of \cref{eq:sigma/Lagrangian} is that the potential $V(h)$ is a function of the heavy field only. This makes it possible to derive the EFT as a derivative expansion, as we will see shortly. Concretely, we will work out the universal expressions for the tree-level EFT Lagrangian up to $\mathcal{O}(\partial^6)$ and the one-loop EFT Lagrangian up to $\mathcal{O}(\partial^4)$.

\subsection*{Tree-Level Matching}

We begin with the tree-level matching, for which we solve the EOM for the heavy field $h$:
\begin{equation}
h_\c = -\frac{1}{M^2}\left[(\pd^2 h) - F F_{,h}\, \hat{g}_{ab}\, (\pd_\mu \pi^a)(\pd^\mu \pi^b) + \frac12\, \bV_{,hhh}\, h^2 + \,\cdots \right] \biggr|_{h=h_\c} \,,
\label{eq:hc}
\end{equation}
where ``$+\cdots$'' denotes higher order terms in $h$. Equation~\eqref{eq:hc} can be solved iteratively to obtain
\begin{align}
h_\c &= \frac{1}{M^2}\, \bF_{,h}\, \hat{g}
- \frac{1}{M^4}\, \bF_{,h}\, (\pd^2 \hat{g})
\notag\\[5pt]
&\quad
+ \frac{1}{M^6}\, \bF_{,h} \left[ M^2 (\bF_{,h}^2 + \bF_{,hh}) - \frac12\, \bF_{,h}\, \bV_{,hhh} \right] \hat{g}^2
+ \mathcal{O}(\partial^6) \,,
\label{eq:sigma/HeavyClassicalEquationOfMotion}
\end{align}
where
\begin{align}
\hat{g} \equiv \eta^{\mu\nu}\hat{g}_{\mu\nu} \qquad \text{with} \qquad \hat{g}_{\mu\nu} \equiv \hat{g}_{ab}(\pi) (\partial_\mu \pi^a) (\partial_\nu \pi^b) \,.
\label{eq:ghat}
\end{align}
Note that $\hat{g}$ starts at $\mathcal{O}(\partial^2)$, and so does $h_\c$; that $h_\c$ carries a positive derivative power is a direct consequence of the absence of nonderivative interactions between $h$ and $\pi^a$. Substituting Eq.~\eqref{eq:sigma/HeavyClassicalEquationOfMotion} into Eq.~\eqref{eq:sigma/Lagrangian}, we obtain the tree-level EFT Lagrangian:
\begin{align}
{\cal L}_{\text{EFT}}^{[0]}[\pi] = {\cal L}[h_\c, \pi] = \,&\, \frac{1}{2}\, \hat{g} + \frac{\bF_{,h}^2}{2M^2}\, \hat{g}^2
+ \frac{\bF_{,h}^2}{2M^4}(\pd_\mu \hat{g})^2 \notag\\[5pt]
& \quad
+ \frac{\bF_{,h}^2}{2M^6}\left[M^2(\bF_{,h}^2 + \bF_{,hh}) - \frac{1}{3}\bF_{,h}\bV_{,hhh}\right]\hat{g}^3 + \mathcal{O}(\partial^8) \,.
\label{eq:sigma/tree_result}
\end{align}

\subsection*{One-Loop Matching}

Next, to obtain the one-loop EFT Lagrangian, we need to derive the matrices entering \cref{eq:geo/eft_lag_expand} from the UV Lagrangian \cref{eq:sigma/Lagrangian}. We find
\begin{subequations}
\begin{align}
\U^h{}_h &= -F F_{,hh}\, \hat{g} + V_{,hh} - M^2 
\,, \\[8pt]
\U^h{}_b &= FF_{,hh} \,\hat{g}_{bc}\, (\pd^\mu \pi^c)(\pd_\mu h)
\,,\\[8pt]
\U^a{}_h &= \frac{F_{,hh}}{F}\, (\pd^\mu \pi^a)(\pd_\mu h) 
\,,\\[8pt]
\U^a{}_b &= \hat{R}^a{}_{cbd}\, (\pd_\mu \pi^c)(\pd^\mu \pi^d) 
+ F_{,h}^2\, \hat{g}_{bc}\, (\pd_\mu \pi^a)(\pd^\mu \pi^c)
\notag\\[3pt]
&\quad
+ \left[ -F_{,h}^2\, \hat{g} - \frac{F_{,hh}}{F}\, (\pd_\mu h)^2 + \frac{F_{,h}}{F}\, V_{,h}\right]\delta^a_b
\,,\\[8pt]
(\Pd_\mu)^h{}_h &= i\,\pd_\mu \,,\qquad\qquad
(\Pd_\mu)^a{}_b = i\,(\hat{\scr D}_\mu)^a{}_b + i\,\frac{F_{,h}}{F}\, (\pd_\mu h)\, \delta^a_b
\,,\\[8pt]
(\Y_{\mu\nu})^h{}_h &= 0 \,,\qquad\qquad 
(\Y_{\mu\nu})^a{}_b = \hat{R}^a{}_{bcd}\, (\pd_\mu \pi^c)(\pd_\nu \pi^d)
\,,\\[8pt]
(\K_\mu)^h{}_b &= -i\,FF_{,h}\, \hat{g}_{bc}\, (\pd_\mu \pi^c)
\,,\qquad\qquad
(\K_\mu)^a{}_h = i\, \frac{F_{,h}}{F}\, (\pd_\mu \pi^a)
\,,
\end{align}
\end{subequations}
where $\hat {\scr D}_\mu$, $\hat R^a{}_{bcd}$ are the covariant derivative and Riemann curvature associated with the connection $\hat \Gamma^a_{bc}$ of the EFT submanifold defined in Eq.~\eqref{eq:sigma_connection}. When $h$ is set to $h_\c$, the minimum derivative orders of these components are given by
\begin{align}
	\U^i{}_j\sim 
	\begin{pmatrix}
	\partial^2 & \partial^4 \\
	\partial^4 & \partial^2
	\end{pmatrix} \,,\qquad
	(\K_\mu)^i{}_j \sim
	\begin{pmatrix}
	 & \partial^1\\
	\partial^1 & 
	\end{pmatrix} \,,\qquad
	(\Y_{\mu\nu})^i{}_j \sim
	\begin{pmatrix}
	\,0 \,&  \\
	 & \partial^2
	\end{pmatrix} \,,
\label{eq:sigma/PowerCountingOfGeometricQuantities}
\end{align}
where empty entries are identically zero by definition, while ``0'' means the entry vanishes in the specific class of UV theories considered here. Importantly, all nonzero entries of the $\U, \K, \Y$ matrices carry positive derivative powers, so only a finite number of terms in the expansion of \cref{eq:geo/eft_lag_expand} need to be evaluated to obtain EFT operators up to any given derivative order. Since $(\Y_{\mu\nu})^h{}_h = 0$, the first line of \cref{eq:geo/eft_lag_expand} coming from log-type traces does not contribute in the present case. For the power-type traces, we keep terms in the expansion in the last two lines of Eq.~\eqref{eq:geo/eft_lag_expand} up to those containing operators of order $\mathcal{O}(\partial^4)$; see App.~\ref{app:sigma} for details. The result is
\begin{align}
{\cal L}_{\text{EFT}}^{[1]} &=
\frac{1}{16\pi^2} \bigg\{
\bigg[ \tfrac{1-L}{2}\, (\bF_{,h}\bV_{,hhh} - M^2\bF_{,hh})
+ \tfrac14\, M^2\, \bF_{,h}^2 \bigg]\, \hat{g}
\notag\\[5pt]
&\quad
+ \bigg[ - \Big( \tfrac{4-3L}{36}\, \bF_{,h}^4 
+ \tfrac{1-2L}{4}\, \bF_{,h}^2\, \bF_{,hh}
+ \tfrac{1-L}{2}\, \bF_{,h}\, \bF_{,hhh}
+ \tfrac{L}{4}\, \bF_{,hh}^2 \Big)
\notag\\[0pt]
&\quad\qquad
+ \tfrac{1}{M^2} \Big( \tfrac{3-2L}{4}\, \bF_{,h}^3\, \bV_{,hhh}
+ \tfrac{1-L}{4}\, \bF_{,h}^2\, \bV_{,hhhh}
+ \tfrac12\, \bF_{,h}\, \bF_{,hh}\, \bV_{,hhh} \Big)
- \tfrac{1}{M^4}\, \Big( \tfrac14\, \bF_{,h}^2\, \bV_{,hhh}^2 \Big) \bigg]\, \hat{g}^2
\notag\\[8pt]
&\quad 
+ \tfrac{5-6L}{18}\, \bF_{,h}^4\, \hat{g}_{\mu\nu}\, \hat{g}^{\mu\nu}
+ \tfrac{4-3L}{18}\, \bF_{,h}^2\, \hat{g}_{ab}\,
(\hat{\scr D}_\mu \pd^\mu \pi)^a (\hat{\scr D}_\nu \pd^\nu \pi)^b
- \tfrac{5-6L}{36}\, \bF_{,h}^2\, \hat{g}_{ab}\,
(\hat{\scr D}_\mu \pd_\nu \pi)^a (\hat{\scr D}^\mu \pd^\nu \pi)^b
\notag\\[8pt]
&\quad
+ \tfrac{49-30L}{36}\, \bF_{,h}^2\, \hat{R}_{abcd}\,
(\pd_\mu \pi^a) (\pd_\nu \pi^b) (\pd^\mu \pi^c) (\pd^\nu \pi^d) \bigg\}
+ \mathcal{O}(\partial^6) \,,
\label{eq:sigma/loop_result}
\end{align}
where, as mentioned before, $L \equiv \log(M^2/\mu^2)$ and we use $\overline{\text{MS}}$ scheme throughout this work.

Equation~\eqref{eq:sigma/loop_result} is the UOLEA for $O(N)$ sigma models, \ie\ the universal one-loop matching formula for any UV theory of the form Eq.~\eqref{eq:sigma/Lagrangian}. It contains 1 operator at $\mathcal{O}(\partial^2)$ and 5 operators at $\mathcal{O}(\partial^4)$. These operators, as well as the operators in the tree-level result \cref{eq:sigma/tree_result}, are all built from tensors on the EFT submanifold, including the vector $(\pd_\mu\pi^a)$ and its covariant derivatives, the metric $\hat g_{ab}$ and the Riemann curvature $\hat R_{abcd}$.

\subsection*{Example: Linear Sigma Model}

To demonstrate the UOLEA at work, let us consider the $O(N)$ linear sigma model as an explicit example. The UV Lagrangian can be written in terms of the $N$-component scalar field $\Phi^i$:
\begin{align}
{\cal L} = \frac12 (\pd_\mu {\bf\Phi}) \cdot (\pd^\mu {\bf\Phi}) - V({\bf\Phi} \cdot {\bf\Phi}) \,,
\end{align}
and we assume the potential leads to spontaneous symmetry breaking $O(N) \to O(N-1)$:
\begin{align}
V({\bf\Phi} \cdot {\bf\Phi}) = -\frac{1}{2}\,\mu^2\, {\bf\Phi} \cdot {\bf\Phi} + \frac{1}{4}\,\lambda\, ({\bf\Phi} \cdot {\bf\Phi})^2 \,.
\end{align}

To cast the Lagrangian in the form of Eq.~\eqref{eq:sigma/Lagrangian}, we use the exponential parameterization as in CCWZ \cite{Coleman:1969sm, Callan:1969sn}:
\begin{align}
{\bf\Phi}(x) = [v + h(x)]\, \xi(x)\, \boldsymbol\chi_0 \,,
\end{align}
where
\begin{align}
\xi(x) = e^{i \pi^a(x) X^a/v} \,,\qquad
{\boldsymbol\chi}_0 = (0, \cdots, 0, 1)^T \,,\qquad
v = \sqrt{\mu^2/\lambda} \,,
\end{align}
and $X^a$ are the (broken) generators of $O(N)$:
\begin{align}
X^a = -i \left(
\begin{array}{ccccc}
0 & & \cdots &  & 0 \\
\vdots & & \vdots & & \vdots \\
0 & & \cdots & & 1 \\
\vdots & & \vdots & & \vdots \\
0 & \cdots & -1 & \cdots & 0 \\
\end{array}\right) \,,\qquad
a = 1 \;,\; \cdots \;,\; N-1 \,.
\end{align}
In the $\{h, \pi^a\}$ coordinates, the Lagrangian reads
\begin{align}
{\cal L} = \frac{1}{2}(\pd_\mu h)^2 + \frac{1}{2} \left(1 + \frac{h}{v}\right)^2 \hat{g}_{ab}(\pi)(\pd_\mu \pi^a)(\pd^\mu \pi^b) - \frac{1}{4}\lambda(h^2 + 2hv)^2 \,,
\label{eq:lsm_lag}
\end{align}
where
\begin{equation}
\hat{g}_{ab}(\pi) = \frac{\sin^2(|\pi|/v)}{|\pi|^2/v^2} \left(\delta^{ab} - \frac{\pi^a\pi^b}{|\pi|^2}\right) + \frac{\pi^a \pi^b}{|\pi|^2} \,,\qquad 
|\pi| = \sqrt{\pi^a\pi^a} \,.
\label{eq:sigma/metric_on_sphere}
\end{equation}
From \cref{eq:lsm_lag} we can readily read off:
\begin{align}
\bF_{,h} = \frac{1}{v} \,,\qquad
\bF_{,hh} = \bF_{,hhh} = 0 \,,\qquad
M^2 = 2\lambda v^2 \,,\qquad
\bV_{,hhh} = 6\lambda v \,.
\end{align}
Substituting these expressions into \cref{eq:sigma/tree_result,eq:sigma/loop_result}, we find the EFT Lagrangian up to one-loop level:
\begin{align}\label{eq:sigma/eft_lag_sph}
{\cal L}_{\text{EFT}} &= 
\frac{1}{2}\left[1 + \frac{\lambda}{16\pi^2}(7-6L)\right]\hat{g} - \frac{\lambda}{16\pi^2}\frac{1}{18M^2}(5-6L)\hat{g}_{ab}(\hat{\scr D}_\mu \pd_\nu \pi)^a(\hat{\scr D}^\mu \pd^\nu \pi)^b \notag\\[5pt]
&\hspace{-10pt}
+ \frac{\lambda}{M^4}\left[1 + \frac{\lambda}{16\pi^2}(8-12L)\right]\hat{g}^2 - \frac{\lambda}{16\pi^2}\frac{\lambda}{3M^4}(13-6L) \,\hat{g}_{\mu\nu}\hat{g}^{\mu\nu} 
+ \mathcal{O}(\partial^6) \,,
\end{align}
where we have used $\hat{R}_{abcd} = \bF_{,h}^2(\hat{g}_{ac}\hat{g}_{bd} - \hat{g}_{ad}\hat{g}_{bc})$ (reflecting the fact that the EFT submanifold is maximally symmetric) to eliminate the term involving $\hat R_{abcd}$. We have also removed terms containing $(\pd^2\pi^a)$ by field redefinitions.

As a cross check, we also performed the matching calculation in the linear parameterization, $\Phi^a = \pi^a$ ($a = 1 \;,\; \cdots \;,\; N-1$), $\Phi^N = v + \sigma$, in which case the EFT operators are organized by canonical dimension. The result is
\begin{align}\label{eq:sigma/eft_lag}
{\cal L}_{\text{EFT}} &=
\frac12 (\pd_\mu \pi^a) (\pd^\mu \pi^a) + \frac{\lambda}{M^2}\left[1 - \frac{\lambda}{16\pi^2}(7-6L)\right] \pi^a\pi^b (\pd_\mu \pi^a) (\pd^\mu \pi^b)
\notag\\[5pt]
&\quad
+ \frac{2\lambda^2}{M^4}\left[1 - \frac{2\lambda}{16\pi^2}(7-6L)\right] (\pi\cdot\pi) \,\pi^a\pi^b (\pd_\mu \pi^a) (\pd^\mu \pi^b)
\notag\\[5pt]
&\quad
+ \frac{\lambda}{M^4} \left[1 - \frac{\lambda}{16\pi^2}\frac{1}{9}(59-6L)\right] [(\pd_\mu \pi^a) (\pd^\mu \pi^a)]^2
\notag\\[5pt]
&\quad
- \frac{\lambda}{16\pi^2}\frac{2\lambda}{9M^4}(17-6L) [(\pd_\mu \pi)\cdot(\pd_\nu \pi)]^2 + \mathcal{O}\bigl(M^{-6}\bigr) \,.
\end{align}
Expanding Eq.~\eqref{eq:sigma/eft_lag_sph} up to dimension-8 and reducing the operators to a minimal basis via field redefinitions gives the same result as in Eq.~\eqref{eq:sigma/eft_lag}. However, the geometric matching result Eq.~\eqref{eq:sigma/eft_lag_sph} includes an infinite series of higher-dimensional operators with up to 4 derivatives, which are packaged into geometric objects on the EFT submanifold.

\section{UOLEA for General Two-derivative Theories}
\label{sec:guolea}

We next turn to the derivation of universal matching formulae for general UV theories containing up to two derivatives:
\begin{equation}
{\cal L} = \frac12\, g_{ij}(\varphi) (\pd_\mu \varphi^i) (\pd^\mu \varphi^j) - V(\varphi) \,.
\label{eqn:LagTwoDerivative}
\end{equation}
Without loss of generality, we assume the fields are canonically normalized and there is no kinetic mixing, so $g_{ij}(\varphi) = \delta_{ij} + \cdots$. Unlike the case of sigma models discussed in the previous section, it is generally not possible to obtain all operators at a given derivative order by computing a finite number of terms, since one can insert an arbitrary number of interaction vertices without increasing the number of derivatives. Instead, we will organize the EFT operators by their canonical dimensions and work out the UOLEA operators that are sufficient for EFT matching up to dimension 6.

A few comments are in order before we proceed with the calculation. Since we are interested in theories with a nontrivial metric, the UV Lagrangian contains irrelevant operators. In the case of sigma models discussed in the previous section, light fields are massless and irrelevant operators are suppressed by the heavy mass scale $M$. More generally, however, a theory may also contain a light mass scale $m$ which may appear in the denominator in some operator basis; for example, adopting the nonlinear parameterization for the Standard Model Higgs field results in powers of $(h/v)$ with $v \ll M$, the mass scale of heavy new particles being integrated out. We allow for such a possibility, but assume that light masses can only appear in the denominator of operators made of light fields, whereas irrelevant operators involving heavy fields must be suppressed by powers of $M$.

Recall that we use square brackets to denote the minimum operator dimension, \eg\ $[\phi] = 1$, $[g_{ij}] = 0$. When determining the truncation of EFT operators, we count light masses as carrying operator dimension one, $[m] = 1$, so for example, both $m^2\phi^2$ and $\phi^4$ are counted as dimension-4, while $\frac{m^2}{M^2}\phi^4$ is counted as dimension-6 (heavy masses do not carry operator dimension, $[M]=0$). We assume that in the UV theory, operators involving only light fields are of dimension 4 or higher, \eg\ there are no terms like $M\phi^3$ or $\frac{M^2}{m^4}\phi^6$ with positive powers of $M$. Meanwhile, $[\Phi] \equiv [\Phi_c(\phi)]$ as determined by the EOM solution is always 2 or higher, which means operators involving heavy fields are also of dimension 4 or higher. Under these assumptions, we summarize in \cref{tab:MinOpdimTwoDerivative} the minimum operator dimensions of various quantities for a UV theory of the form of \cref{eqn:LagTwoDerivative}. Note that while components of the connection and curvature tensor may have negative operator dimensions due to the possible appearance of $m$ in the denominator, all components of the $\U, \Pd, \K, \Y$ matrices have positive minimum operator dimensions.

\begin{table}[t]
\renewcommand{\arraystretch}{1.0}
\setlength{\arrayrulewidth}{.2mm}
\setlength{\tabcolsep}{1em}
\centering
\begin{tabular}{ccccc}
\toprule
$[g_{ij}] \,,\, [g^{ij}] $ & $[\Gamma^H_{ij}]$ & $[\Gamma^L_{ij}]$ & $[\W_{,i}]$ & $[\W_{,ij}]$ \\[2pt]
\midrule
$\mqty(\,0\, & 1\, \\ \,1\, & 0\, )$ & $\mqty(\,0\, & 0\, \\ \,0\, & 0\, )$ & $\mqty(\,0 & 0 \\ \,0 & -1 )$ & $\mqty(2 \\ 3)$ & $\mqty(\,1\, & 1\, \\ \,1\, & 2\,)$ \\
\midrule\midrule
& $[R^{H}{}_{Hij}]$ & $[R^{H}{}_{Lij}] \,,\, [R^{L}{}_{Hij}]$ & $[R^{L}{}_{Lij}]$ & $[\Phi_\c]$ \\[2pt]
\midrule
& $\mqty(\,0\, & 0\, \\ \,0\, & 0\, )$ & $\mqty(0 & -1 \\ -1 & -1 )$ & $\mqty(0 & -1 \\ -1 & -2 )$ & 2 \\
\midrule\midrule
{\small General case results:} & $[\,\U^i{}_j]$ & $[\Pd^i{}_j]$ & $[\K^i{}_j]$ & $[\,\Y^i{}_j]$ \\[1pt]
\midrule
& $\mqty(\,1\, & 1\, \\ \,1\, & 2\,)$ & $\mqty(\,1\, & \\ & 1\,)$ & $\mqty( & \,2\, \\ 2\, & )$ & $\mqty(\,4\, & \\ & 2\,)$ \\
\bottomrule
\end{tabular}
\caption{\label{tab:MinOpdimTwoDerivative}
Minimum operation dimensions of various quantities for a general UV theory containing up to two derivatives, \cref{eqn:LagTwoDerivative}. Empty entries mean the matrix blocks are zero by definition. The minimum operator dimensions of $\Phi_\c$ and some entries of the $\U$ matrix may be higher under certain assumptions about the UV theory; see \cref{eq:Udim_Z2,eq:mindim_z2_noPhi}.}
\end{table}

In addition to the general case where the minimum operator dimensions are summarized in \cref{tab:MinOpdimTwoDerivative}, it is also interesting to consider a more restricted scenario where the UV theory has a $\mathbb{Z}_2$ symmetry under which the light fields change sign, $\phi\to-\phi$. In this case, some components of $\Gamma^i_{jk}$, $R^i{}_{jkl}$ and $\W_{,ij}$ start at higher operator dimensions, resulting in a higher minimum operator dimension for the heavy-heavy block of the $\U$ matrix:
\begin{equation}
[\,\U^i{}_j] = \left(
\begin{array}{cc}
2 & 1 \\
1 & 2
\end{array}
\right) \qquad
(\mathbb{Z}_2 \text{ symmetric case}) \,,
\label{eq:Udim_Z2}
\end{equation}
while the minimum operator dimensions of the $\Pd$, $\K$ and $\Y$ matrices are the same as in the general case.

We may further consider the possibility that the heavy fields $\Phi$ do not have nonderivative interactions, in which case $\W(\varphi) = \W(\phi)$ in \cref{eq:V_sep} is a function of the light fields only. As we will discuss in the next section, it is always possible to eliminate nonderivative interactions of the heavy fields in favor of derivative interactions via (nonderivative) field redefinitions. The advantage of such an operator basis is that $\Phi_\c$ and the heavy-light block of the $\U$ matrix start at higher operator dimensions:
\begin{align}
[\Phi_\c] =4 \,,\qquad
[\,\U^i{}_j] = \left(
\begin{array}{cc}
2 & 3 \\
1 & 2
\end{array}
\right) \qquad
(\mathbb{Z}_2 \text{ symmetric case},\; \W(\varphi) = \W(\phi)) \,,
\label{eq:mindim_z2_noPhi}
\end{align}
thereby further reducing the number of terms of UOLEA truncated at dimension six.

After enumerating the terms in \cref{eq:geo/eft_lag_expand} up to minimum operator dimension 6, we can proceed to evaluate these terms. Each term is the trace (over internal indices) of a product of matrices $(q^2 - {\bf M}^2)^{-1}$, $\U$, $\Pd_\mu$, $\K_\mu$ and $\Y_{\mu\nu}$. Since the propagator $(q^2 - {\bf M}^2)^{-1}$ is block-diagonal, once all matrices are written out in terms of the heavy and light blocks, the momentum integral factors out and can be evaluated using standard methods. As usual we adopt dimensional regularization and $\overline{\text{MS}}$ scheme. The remaining part is a trace of a product of blocks of the $\U$, $\Pd$, $\K$, $\Y$ matrices, and is manifestly covariant on the EFT submanifold.

Let us demonstrate this procedure with an example:
\begin{equation}
T_{UUK} \equiv \int\frac{\md^dq}{(2\pi)^d} \,\op{tr} \biggl[ \tfrac{1}{q^2 - {\bf M}^2} \, \widetilde \U \, \tfrac{1}{q^2 - {\bf M}^2} \, \widetilde \U \, \tfrac{1}{q^2 - {\bf M}^2} \,\bigl\{ q_\mu - \Yt_{\mu\nu}\,\partial_q^\nu, \widetilde \K^\mu \bigr\}\biggr] \,,
\end{equation}
where it is implicit that we replace $\Phi$ by $\Phi_\c[\phi]$. Assuming the $\mathbb{Z}_2$ symmetric case with $W(\varphi) = W(\phi)$ for simplicity, we have $[\Pd] = [\,\U_{LH}] = 1$, $[\,\U_{HH}] = [\,\U_{LL}] = [\K_{HL}] = [\K_{LH}] = [\,\Y_{LL}] = 2$, $[\,\U_{HL}] = 3$ and $[\,\Y_{HH}] = 4$; see \cref{tab:MinOpdimTwoDerivative,eq:mindim_z2_noPhi}. Substituting in the definitions of $\,\widetilde{\U}$, $\widetilde{\K}_\mu$ and $\Yt_{\mu\nu}$ from \cref{eq:YUKtilde} and dropping terms with odd powers of momenta (which integrate to zero), we obtain
\begin{align}
T_{UUK} &= \int\frac{\md^dq}{(2\pi)^d} \,\op{tr} \biggl( \tfrac{1}{q^2 - {\bf M}^2}\, \U\, \tfrac{1}{q^2 - {\bf M}^2}\, \U\, \tfrac{1}{q^2 - {\bf M}^2}\, [\Pd_\mu, \K^\mu] \biggr)
\notag\\[5pt]
&\quad
+ \int\frac{\md^dq}{(2\pi)^d}\, \op{tr} \biggl( \tfrac{1}{q^2 - {\bf M}^2}\, \U\, \tfrac{1}{q^2 - {\bf M}^2}\, [\Pd_\nu ,\U]\, \pd_q^\nu\, \tfrac{1}{q^2 - {\bf M}^2}\, 2 q_\mu \K^\mu \biggr)
\notag\\[5pt]
&\quad
+ \int\frac{\md^dq}{(2\pi)^d}\, \op{tr} \biggl( \tfrac{1}{q^2 - {\bf M}^2}\, [\Pd_\nu , \U]\,\pd_q^\nu\, \tfrac{1}{q^2 - {\bf M}^2}\, \U\, \tfrac{1}{q^2 - {\bf M}^2}\, 2 q_\mu \K^\mu \biggr)
\,+ \cdots \,,
\label{eq:T_UUK}
\end{align}
where additional terms in ``$+\cdots$'' do not contribute to operators up to dimension 6. Next, we write out the blocks of the matrices and factor out the momentum integral. Taking the first term in Eq.~\eqref{eq:T_UUK} for example, we find
\begin{align}
T_{UUK,1} &\equiv
\int\frac{\md^dq}{(2\pi)^d} \,\op{tr} \biggl( \tfrac{1}{q^2 - {\bf M}^2} \, \U \, \tfrac{1}{q^2 - {\bf M}^2}\, \U\, \tfrac{1}{q^2 - {\bf M}^2} \,[\Pd_\mu , \K^\mu] \biggr)
\notag\\[8pt]
&=
\int\frac{\md^dq}{(2\pi)^d} \, \biggl[ \tfrac{1}{(q^2 - M^2)^2\, q^2}\, \op{tr} \bigl(\, \U_{LH}\, \U_{HH}\, [\Pd_\mu , \K^\mu]_{HL} \bigr)
\notag\\[2pt]
&\hspace{80pt}
+ \tfrac{1}{(q^2 - M^2)\, q^4}\, \op{tr} \bigl(\, \U_{LL}\, \U_{LH}\, [\Pd_\mu , \K^\mu]_{HL} \bigr) \biggr] \,.
\end{align}
Note that $\bigl[\Pd_\mu , \K^\mu\bigr]$ has only off-diagonal blocks, and we have only kept terms up to minimum operator dimension 6. Evaluating the momentum integral, we obtain
\begin{align}
\mathcal{L}_\text{EFT}^{[1]} \supset -\frac{i}{2}\,\frac{1}{3}\, T_{UUK,1} &= \frac{1}{16\pi^2} \frac{1}{6M^2} \biggl[ -\op{tr} \bigl(\, \U_{LH}\, \U_{HH}\, [\Pd_\mu , \K^\mu]_{HL} \bigr) \, \notag\\[5pt]
&\qquad
+ \left( \frac{2}{\epsilon} + 1 - \log\frac{M^2}{\mu^2} \right) \op{tr} \bigl(\, \U_{LL}\, \U_{LH} \, [\Pd_\mu, \K^\mu]_{HL} \bigr)\, \biggr] \,,
\end{align}
where $\mu$ is the renormalization scale and $\epsilon=4-d$.

\begin{table}
	{\small
		\begin{center}
			\begin{tabular}{|c|c|}
				\hline
				\multicolumn{2}{|c|}{ \Tstrut\Bstrut ${\cal O}(\,\U)$ terms} \\
				\hline
				\Tstrut\Bstrut $\dfrac{1}{2}(1-L)M^2$ & $\ca{\U_{HH}}$ \\
				\Hline
				\multicolumn{2}{|c|}{ \Tstrut\Bstrut ${\cal O}(\,\U^2)$ terms} \\
				\hline
				\Tstrut\Bstrut $-\dfrac{1}{4}L$ & $\ca{ (\,\U_{HH})^2}$ \\
				\hline
				\Tstrut\Bstrut $\dfrac{1}{2}(1-L)$ & $\ca{\U_{HL}\,\U_{LH}}$ \\
				\Hline
				\multicolumn{2}{|c|}{ \Tstrut\Bstrut ${\cal O}(\,\U^3)$ terms} \\
				\hline
				\Tstrut\Bstrut $-\dfrac{1}{12M^2}$ & $\ca{ (\,\U_{HH})^3}$ \\
				\hline
				\Tstrut\Bstrut $- \dfrac{1}{2M^2}$ & $\ca{ \U_{HH} \,\U_{HL} \,\U_{LH}}$ \\
				\hline
				\Tstrut\Bstrut $\dfrac{1}{2M^2}(1-L)$ & $\ca{ \U_{HL} \,\U_{LL} \,\U_{LH}}$ \\
				\hline
			\end{tabular}
			\hspace{2pt}
			\begin{tabular}{|c|c|}
				\hline
				\multicolumn{2}{|c|}{ \Tstrut\Bstrut ${\cal O}(\,\U^4)$ terms} \\
				\hline
				\Tstrut\Bstrut $\dfrac{1}{4M^4}$ & $\cb{ (\,\U_{HH})^2 \,\U_{HL} \,\U_{LH}}$ \\
				\hline
				\Tstrut\Bstrut $- \dfrac{1}{2M^4}(2-L)$ & $\cb{ \U_{HH} \,\U_{HL} \,\U_{LL} \,\U_{LH}}$ \\
				\hline
				\Tstrut\Bstrut $- \dfrac{1}{4M^4}(2-L)$ & $\cb{ \U_{HL} \,\U_{LH} \,\U_{HL} \,\U_{LH}}$ \\
				\hline
				\Tstrut\Bstrut $\dfrac{1}{2M^4}(1-L)$ & $\cb{ \U_{HL} \,\U_{LL} \,\U_{LL} \,\U_{LH}}$ \\
				\Hline
				\multicolumn{2}{|c|}{ \Tstrut\Bstrut ${\cal O}(\,\U^5)$ terms} \\
				\hline
				\Tstrut\Bstrut $\dfrac{1}{4M^6}(5-2L)$ & $\cb{ \U_{HH} \,\U_{HL} \,\U_{LH} \,\U_{HL} \,\U_{LH}}$ \\
				\hline
				\Tstrut\Bstrut $- \dfrac{1}{2M^6}(3-2L)$ & $\cb{ \U_{HL} \,\U_{LH} \,\U_{HL} \,\U_{LL} \,\U_{LH}}$ \\
				\Hline
				\multicolumn{2}{|c|}{ \Tstrut\Bstrut ${\cal O}(\,\U^6)$ terms} \\
				\hline
				\Tstrut\Bstrut $\dfrac{1}{12M^8}(11-6L)$ & $\cb{ (\,\U_{HL} \,\U_{LH})^3}$ \\
				\hline
			\end{tabular}
		\end{center}
	}
	\caption{Zero-derivative terms in the UOLEA up to dimension-6 for UV theories where light fields have a $\mathbb{Z}_2$ symmetry $\phi\to-\phi$.}
	\label{tab:guolea/0d_ab}
\end{table}
\begin{table}
	{\small
		\begin{center}
			\begin{tabular}{|c|c|}
				\hline
				\multicolumn{2}{|c|}{ \Tstrut\Bstrut ${\cal O}(\,\U^2\Pd^2)$ terms} \\
				\hline
				\Tstrut\Bstrut $-\dfrac{1}{24M^2}$ & $\ca{[\Pd_\mu ,\U]_{HH} [\Pd^\mu ,\U]_{HH}}$ \\
				\hline
				\Tstrut\Bstrut $-\dfrac{1}{4M^2}$ & $\ca{ [\Pd_\mu ,\U]_{HL} [\Pd_\mu ,\U]_{LH}}$ \\
				\Hline
				\multicolumn{2}{|c|}{ \Tstrut\Bstrut ${\cal O}(\,\U^3\Pd^2)$ terms} \\
				\hline
				\Tstrut\Bstrut $\dfrac{1}{4M^4}$ & $ \cb{\U_{HH} [\Pd_\mu ,\U]_{HL}[\Pd^\mu ,\U]_{LH}}$ \\
				\hline
				\Tstrut\Bstrut $\dfrac{1}{6M^4}$ & $\cb{[\Pd_\mu, \U]_{HH} [\Pd^\mu, \U]_{HL}U_{LH}} \cb{ \;+\; \U_{HL} [\Pd^\mu ,\U]_{LH} [\Pd_\mu ,\U]_{HH}}$ \\
				\hline
				\Tstrut\Bstrut $-\dfrac{1}{4M^4}(5-2L)$ & $\cb{ [\Pd_\mu ,\U ]_{HL}\,\U_{LL} [\Pd^\mu ,\U ]_{LH}}$ \\
				\hline
				\Tstrut\Bstrut $-\dfrac{1}{4M^4}$ & $ \cb{\U_{HL} [\Pd_\mu ,\U]_{LL} [\Pd^\mu ,\U]_{LH}} \cb{ \;+\; [\Pd_\mu ,\U]_{HL} [\Pd^\mu ,\U]_{LL}\,\U_{LH}}$ \\
				\Hline
				\multicolumn{2}{|c|}{ \Tstrut\Bstrut ${\cal O}(\,\U^4\Pd^2)$ terms} \\
				\hline
				\Tstrut\Bstrut $\dfrac{1}{2M^6}$ & $\cb{\U_{HL} \,\U_{LH} [\Pd_\mu ,\U]_{HL} [\Pd^\mu ,\U]_{LH}}$ \\
				\hline
				\Tstrut\Bstrut $\dfrac{1}{12M^6}(17-6L)$ & $\cb{[\Pd_\mu ,\U]_{HL} \,\U_{LH} \,\U_{HL} [\Pd^\mu ,\U]_{LH}}$ \\
				\hline
				\Tstrut\Bstrut $\dfrac{5}{24M^6}$ & $\cb{\U_{HL} [\Pd_\mu ,\U]_{LH} \,\U_{HL} [\Pd^\mu ,\U]_{LH}} \cb{ \;+\;  [\Pd_\mu ,\U]_{HL} \,\U_{LH}[\Pd^\mu ,\U]_{HL} \,\U_{LH}}$ \\
				\Hline
				\multicolumn{2}{|c|}{ \Tstrut\Bstrut ${\cal O}(\K^2)$ terms} \\
				\hline
				\Tstrut\Bstrut $\dfrac{1}{4}M^2$ & $\ca{(\K_\mu \K^\mu)_{HH}}$ \\
				\Hline
				\multicolumn{2}{|c|}{ \Tstrut\Bstrut ${\cal O}(\,\U \K^2)$ terms} \\
				\hline
				\Tstrut\Bstrut $\dfrac{1}{4}$ & $\ca{\U_{HH} (\K_\mu \K^\mu)_{HH}}$ \\
				\hline
				\Tstrut\Bstrut $\dfrac{1}{4}(3-2L)$ & $\ca{\U_{LL} (\K_\mu \K^\mu)_{LL}}$ \\
				\Hline
				\multicolumn{2}{|c|}{ \Tstrut\Bstrut ${\cal O}(\,\U^2\K^2)$ terms} \\
				\hline
				\Tstrut\Bstrut $-\dfrac{1}{2M^2}(2-L)$ & $\cb{\U_{HL}(\K_\mu \K^\mu)_{LL} \,\U_{LH}}$ \\
				\hline
				\Tstrut\Bstrut $-\dfrac{1}{4M^2}$ & $\ca{\U_{LH} (\K_\mu)_{HL} \,\U_{LH} (\K^\mu)_{HL}} \cb{\;+\; \U_{HL} (\K_\mu)_{LH} \,\U_{HL} (\K^\mu)_{LH}}$ \\
				\Hline
				\multicolumn{2}{|c|}{ \Tstrut\Bstrut ${\cal O}(\,\U \K\Pd)$ terms} \\
				\hline
				\Tstrut\Bstrut $\dfrac{1}{4}$ & $\ca{\U_{HL} [\Pd_\mu, \K^\mu]_{LH}} \ca{\;-\; [\Pd_\mu, \K^\mu]_{HL}\, \U_{LH}}$ \\
				\Hline
				\multicolumn{2}{|c|}{ \Tstrut\Bstrut ${\cal O}(\,\U^2\K\Pd)$ terms} \\
				\hline
				\Tstrut\Bstrut $-\dfrac{1}{4M^2}$ & $\ca{[\Pd_\mu ,\U]_{HH} (\K^\mu)_{HL} \,\U_{LH}} \cb{\;-\; \U_{HL}(\K^\mu)_{LH}[\Pd_\mu ,\U]_{HH}}$ \\
				\hline
				\Tstrut\Bstrut $\dfrac{1}{4M^2}$ & $\ca{(\K^\mu)_{HL} [\Pd_\mu ,\U]_{LL} \,\U_{LH}} \cb{\;-\; \U_{HL}[\Pd_\mu ,\U]_{LL}(\K^\mu)_{LH}}$ \\
				\hline
				\Tstrut\Bstrut $\dfrac{1}{2M^2}(2-L)$ & $\ca{(\K^\mu)_{HL} \,\U_{LL} [\Pd_\mu ,\U]_{LH}} \cb{\;-\; [\Pd_\mu ,\U]_{HL} \,\U_{LL}(\K^\mu)_{LH}}$ \\
				\Hline
				\multicolumn{2}{|c|}{ \Tstrut\Bstrut ${\cal O}(\,\U^3\K\Pd)$ terms} \\
				\hline
				\Tstrut\Bstrut $\dfrac{1}{4M^4}(5-2L)$ & $\cb{[\Pd_\mu ,\U]_{HL} \,\U_{LH} \,\U_{HL}(\K^\mu)_{LH}} \cb{\;-\; \U_{HL}[\Pd_\mu ,\U]_{LH}(\K^\mu)_{HL} \,\U_{LH}}$ \\
				\hline
				\Tstrut\Bstrut $\dfrac{1}{4M^4}$ & $\cb{\U_{HL} \,\U_{LH}[\Pd_\mu ,\U]_{HL}(\K^\mu)_{LH}} \cb{\;-\; \U_{HL} \,\U_{LH}(\K^\mu)_{HL}[\Pd_\mu ,\U]_{LH}}$ \\
				\hline
			\end{tabular}
		\end{center}
	}
	\caption{Two-derivative terms in the UOLEA up to dimension-6 for UV theories where light fields have a $\mathbb{Z}_2$ symmetry $\phi\to-\phi$.
	}
	\label{tab:guolea/2d_ab}
\end{table}
\begin{table}
	{\small
		\begin{center}
			\begin{tabular}{|c|c|}
				\hline
				\multicolumn{2}{|c|}{ \Tstrut\Bstrut ${\cal O}(\K^2{\Y})$ terms} \\
				\hline
				\Tstrut\Bstrut $\dfrac{1}{18}(11-6L)$ & $\ca{(\K^\mu)_{HL}(\Y_{\mu\nu})_{LL} (\K^\nu)_{LH}}$ \\
				\Hline
				\multicolumn{2}{|c|}{ \Tstrut\Bstrut ${\cal O}(\K^2 \Pd^2)$ terms} \\
				\hline
				\Tstrut\Bstrut $-\dfrac{1}{18}(4-3L)$ & $\ca{[\Pd_\mu, \K^\mu]_{HL} [\Pd_\nu, \K^\nu]_{LH}}$ \\
				\hline
				\Tstrut\Bstrut $\dfrac{1}{36}(5-6L)$ & $\ca{[\Pd^\nu, \K^\mu]_{HL} [\Pd_\nu, \K_\mu]_{LH}}$ \\
				\Hline 
				\multicolumn{2}{|c|}{ \Tstrut\Bstrut ${\cal O}(\,\U \K{\Y}\Pd)$ terms} \\
				\hline
				\Tstrut\Bstrut $-\dfrac{1}{36M^2}(17-6L)$ & $\ca{(\K_\nu)_{HL} (\Y^{\mu\nu})_{LL} [\Pd_\mu ,\U]_{LH}} \cb{\;+\; [\Pd_\mu ,\U]_{HL} (\Y^{\mu\nu})_{LL} (\K_\nu)_{LH}}$ \\
				\hline
				\Tstrut\Bstrut $-\dfrac{1}{36M^2}(5-6L)$ & $\ca{[\Pd_\mu, \K_\nu]_{HL} (\Y^{\mu\nu})_{LL} \,\U_{LH}} \cb{\;+\; \U_{HL} (\Y^{\mu\nu})_{LL} [\Pd_\mu, \K_\nu]_{LH}}$ \\
				\Hline
				\multicolumn{2}{|c|}{ \Tstrut\Bstrut ${\cal O}(\,\U \K\Pd^3)$ terms} \\
				\hline
				\Tstrut\Bstrut $-\dfrac{1}{12M^2}$ & $\ca{[\Pd_\mu, \K^\mu]\big._{HL} \bigl[\Pd_\nu, [\Pd^\nu ,\U]\bigr]_{LH}} \cb{\;-\; \bigl[[\Pd_\nu, [\Pd^\nu ,\U]\bigr]_{HL} [\Pd_\mu, \K^\mu]\big._{LH}}$ \\
				\Hline
				\multicolumn{2}{|c|}{ \Tstrut\Bstrut ${\cal O}(\,\U^2{\Y}^2)$ terms} \\
				\hline
				\Tstrut\Bstrut $\dfrac{1}{72M^4}(5-6L)$ & $\cb{\U_{HL}(\scrg_{\mu\nu})_{LL}(\scrg^{\mu\nu})_{LL}\,\U_{LH}}$ \\
				\Hline
				\multicolumn{2}{|c|}{ \Tstrut\Bstrut ${\cal O}(\,\U^2{\Y}\Pd^2)$ terms} \\
				\hline
				\Tstrut\Bstrut $-\dfrac{1}{36M^4}(17-6L)$ & $\cb{[\Pd_\mu ,\U]_{HL}(\scrg^{\mu\nu})_{LL} [\Pd_\nu ,\U]_{LH}}$ \\
				\Hline
				\multicolumn{2}{|c|}{ \Tstrut\Bstrut ${\cal O}(\,\U^2\Pd^4)$ terms} \\
				\hline
				\Tstrut\Bstrut $\dfrac{1}{12M^4}$ & $\cb{\bigl[ \Pd_\mu, [\Pd^\mu ,\U]\bigr]_{HL} \bigl[\Pd_\nu, [\Pd^\nu ,\U]\bigr]_{LH}}$ \\
				\hline
			\end{tabular}
		\end{center}
	}
	\caption{Four-derivative terms in the UOLEA up to dimension-6 for UV theories where light fields have a $\mathbb{Z}_2$ symmetry $\phi\to-\phi$.}
	\label{tab:guolea/4d_ab}
\end{table}
\begin{table}
	{\small
		\begin{center}
			\begin{tabular}{|c|c|}
				\hline
				\multicolumn{2}{|c|}{ \Tstrut\Bstrut ${\cal O}(\,\U^4)$ terms} \\
				\hline
				\Tstrut\Bstrut $\dfrac{1}{48M^4}$ & $\cc{ (\,\U_{HH})^4}$ \\
				\Hline
				\multicolumn{2}{|c|}{ \Tstrut\Bstrut ${\cal O}(\,\U^5)$ terms} \\
				\hline
				\Tstrut\Bstrut $- \dfrac{1}{120M^6}$ & $\cc{ (\,\U_{HH})^5}$ \\
				\hline
				\Tstrut\Bstrut $- \dfrac{1}{6M^6}$ & $\cc{ (\,\U_{HH})^3 \,\U_{HL} \,\U_{LH}}$ \\
				\hline
				\Tstrut\Bstrut $\dfrac{1}{4M^6}(5-2L)$ & $\cc{ (\,\U_{HH})^2 \,\U_{HL} \,\U_{LL} \,\U_{LH}}$ \\
				\hline
			\end{tabular}
			\hspace{2pt}
			\begin{tabular}{|c|c|}
				\hline
				\multicolumn{2}{|c|}{ \Tstrut\Bstrut ${\cal O}(\,\U^6)$ terms} \\
				\hline 
				\Tstrut\Bstrut $\dfrac{1}{240M^8}$ & $\cc{ (\,\U_{HH})^6}$ \\
				\hline
				\Tstrut\Bstrut $\dfrac{1}{8M^8}$ & $\cc{ (\,\U_{HH})^4 \,\U_{HL} \,\U_{LH}}$ \\
				\hline
				\Tstrut\Bstrut $- \dfrac{1}{12M^8}(17-6L)$ & $\cc{ (\,\U_{HH})^2 (\,\U_{HL} \,\U_{LH})^2}$ \\
				\hline
				\Tstrut\Bstrut $- \dfrac{1}{24M^8}(17-6L)$ & $\cc{ (\,\U_{HH} \,\U_{HL} \,\U_{LH})^2}$ \\
				\hline
			\end{tabular}
		\end{center}
		\begin{center}
			\begin{tabular}{|c|c|}
				\hline
				\multicolumn{2}{|c|}{ \Tstrut\Bstrut ${\cal O}(\,\U^3 \Pd^2)$ terms} \\
				\hline
				\Tstrut\Bstrut $\dfrac{1}{24M^4}$ & $\cc{ \U_{HH}[\Pd_\mu, \U]_{HH}[\Pd^\mu, \U]_{HH}}$ \\ 
				\Hline
				\multicolumn{2}{|c|}{ \Tstrut\Bstrut ${\cal O}(\,\U^4 \Pd^2)$ terms} \\
				\hline
				\Tstrut\Bstrut $-\dfrac{1}{40M^6}$ & $\cc{ (\,\U_{HH})^2[\Pd_\mu, \U]_{HH}[\Pd^\mu ,\U]_{HH}}$ \\
				\hline
				\Tstrut\Bstrut $-\dfrac{1}{60M^6}$ & $\cc{ \U_{HH}[\Pd_\mu ,\U]_{HH}\,\U_{HH}[\Pd^\mu ,\U]_{HH}}$ \\
				\hline
				\Tstrut\Bstrut $-\dfrac{1}{4M^6}$ & $\cc{(\,\U_{HH})^2[\Pd_\mu ,\U]_{HL}[\Pd^\mu ,\U]_{LH}}$ \\
				\hline
				\Tstrut\Bstrut $-\dfrac{5}{24M^6}$ & $\cc{\U_{HH}[\Pd_\mu ,\U]_{HH}\,\U_{HL}[\Pd^\mu ,\U]_{LH}} \cc{ \;+ \; [\Pd_\mu ,\U]_{HH}\,\U_{HH}[\Pd^\mu ,\U]_{HL}\,\U_{LH}}$ \\
				\hline
				\Tstrut\Bstrut $-\dfrac{1}{8M^6}$ & $\cc{[\Pd_\mu ,\U]_{HH}[\Pd^\mu ,\U]_{HH}\,\U_{HL}\,\U_{LH}}$ \\
				\hline
				\Tstrut\Bstrut $-\dfrac{1}{8M^6}$ & $\cc{ \U_{HH}[\Pd_\mu ,\U]_{HH}[\Pd^\mu ,\U]_{HL}\,\U_{LH}} \cc{\; +\; [\Pd_\mu ,\U]_{HH}\,\U_{HH}\,\U_{HL}[\Pd^\mu ,\U]_{LH}}$ \\
				\Hline
				\multicolumn{2}{|c|}{ \Tstrut\Bstrut ${\cal O}(\,\U^3\K\Pd)$ terms} \\
				\hline
				\Tstrut\Bstrut $-\dfrac{1}{12M^4}$ & $\cc{\U_{HH}[\Pd_\mu ,\U]_{HH}\,\U_{HL}(\K^\mu)_{LH}} \cc{\;-\; [\Pd_\mu ,\U]_{HH}\,\U_{HH}(\K^\mu)_{HL}\,\U_{LH}}$ \\
				\hline
				\Tstrut\Bstrut $-\dfrac{1}{6M^4}$ & $\cc{[\Pd_\mu ,\U]_{HH}\,\U_{HH}\,\U_{HL}(\K^\mu)_{LH}} \cc{\;-\; \U_{HH}[\Pd_\mu ,\U]_{HH}(\K^\mu)_{HL}\,\U_{LH}}$ \\
				\Hline
				\multicolumn{2}{|c|}{ \Tstrut\Bstrut ${\cal O}(\,\U^2\Pd^4)$ terms} \\
				\hline
				\Tstrut\Bstrut $\dfrac{1}{240M^4}$ & $\cc{\bigl[\Pd_\mu, [\Pd^\mu, \U]\bigr]_{HH} \bigl[\Pd_\nu, [\Pd^\nu, \U]\bigr]_{HH}}$ \\
				\hline
			\end{tabular}
	\end{center}}
	\caption{Additional terms in the UOLEA up to dimension-6 for UV theories without a $\mathbb{Z}_2$ symmetry.}
	\label{tab:guolea/0d_c}
\end{table}

Following the same procedure we can enumerate and evaluate all the terms in the UOLEA up to dimension-6, aided by the \texttt{STrEAM}~\cite{Cohen:2020qvb} and \texttt{SuperTracer}~\cite{Fuentes-Martin:2020udw} packages. The result takes the form:
\begin{align}
\mathcal{L}_\text{EFT}^{[1]} = \frac{1}{16\pi^2}\, \sum_i c_i \op{tr}\mathcal{O}_i \;,
\end{align}
where $c_i$ and $\mathcal{O}_i$ are listed in the first and second columns of Tables~\ref{tab:guolea/0d_ab}-\ref{tab:guolea/0d_c}, respectively. As in \cref{sec:geo/action,sec:sigma}, we abbreviate $\log(M^2/\mu^2) \equiv L$ in the operator coefficients. In the case of a $\mathbb{Z}_2$-symmetric UV theory, only Tables~\ref{tab:guolea/0d_ab}-\ref{tab:guolea/4d_ab} are needed to obtain EFT operators up to dimension 6. We further \ca{highlight} those operators that are dimension-6 or lower in the special case discussed around \cref{eq:mindim_z2_noPhi} where heavy fields do not have nonderivative interactions, $\W(\varphi) = \W (\phi)$; only these terms will be needed in the example in Sec.~\ref{sec:singlet}. The UOLEA operators in Tables~\ref{tab:guolea/0d_ab}-\ref{tab:guolea/0d_c} are written in terms of $\,\U$, $\Pd_\mu$, $\K_\mu$ and $\Y_{\mu\nu}$, whose definitions we reproduce here for convenience:
\begin{subequations}
\begin{align}
\U^i{}_j &= R\indices{^i_{kjl}} \bigl(\partial_\mu \varphi^k\bigr) \bigl(\partial^\mu \varphi^l\bigr) + \W\indices{_;^i_{\;j}}
+ \Bigl[ \bigl( g^{iA}-\delta^{iA} \bigr) \delta_{Aj} - g^{ik} \Gamma^A_{jk} \delta_{AB} \Phi^B \Bigr] M^2 \,, \\[8pt]
(\Pd_\mu)^A{}_B &= i\, \delta^A_B \pd_\mu + i\, (\pd_\mu \varphi^k) \,\Gamma^A_{kB}
\,, \qquad
(\Pd_\mu)^a{}_b = i\, \delta^a_b \pd_\mu + i\, (\pd_\mu \varphi^k) \,\Gamma^a_{kb} \,, \\[8pt]
(\K_\mu)^A{}_b &= i\, (\pd_\mu \varphi^k) \,\Gamma^A_{kb}
\,, \qquad
(\K_\mu)^a{}_B = i\, (\pd_\mu \varphi^k) \,\Gamma^a_{kB} \,,\\[8pt]
\hspace{-20pt}
(\Y_{\mu\nu})^A{}_B &= -[\Pd_\mu, \Pd_\nu]^A{}_B = (R^{A}{}_{Bkl} - \Gamma^A_{kc}\Gamma^c_{lB} + \Gamma^A_{lc}\Gamma^c_{kB})(\pd_\mu \varphi^k) (\pd_\mu \varphi^l) \,, \\[8pt]
(\Y_{\mu\nu})^a{}_b &= -[\Pd_\mu, \Pd_\nu]^a{}_b = (R^{a}{}_{bkl} - \Gamma^a_{kC}\Gamma^C_{lb} + \Gamma^a_{lC}\Gamma^C_{kb})(\pd_\mu \varphi^k) (\pd_\mu \varphi^l) \,,
\end{align}
\end{subequations}
with all other blocks vanishing.

\section{Example: Singlet Scalar Extended Standard Model}
\label{sec:singlet}

As an example application of the geometric UOLEA derived in the previous section, let us consider the Standard Model (SM) extended by a singlet scalar $S$:
\begin{align}
\mathcal{L} = \mathcal{L}_\text{SM} + \frac12 (\pd_\mu S)^2 - \frac12 M^2 S^2 - A|H|^2S - \frac12 \kappa|H|^2 S^2 - \frac{1}{3!} \mu_S S^3 - \frac{1}{4!} \lambda_S S^4 \,,
\end{align}
where $H$ is the Standard Model Higgs doublet, and we assume the dimensionful couplings $A, \mu_S \sim\mathcal{O}(M)$. The full matching result of this model onto SMEFT up to dimension-6 and one-loop level has been obtained using Feynman diagrams~\cite{Haisch:2020ahr}, functional methods~\cite{Cohen:2020fcu}, and a combination of both~\cite{Jiang:2018pbd}. Here we focus on reproducing the scalar sector contributions for the purpose of demonstrating the geometric UOLEA at work. Concretely, we turn off gauge and Yukawa couplings, so the only relevant terms in the UV Lagrangian are
\begin{align}
\mathcal{L} = |\pd_\mu H|^2 + \frac{1}{2}(\pd_\mu S)^2 - V(S, H) \,,
\label{eq:L_singlet}
\end{align}
where the scalar potential is given by
\begin{align}
V(S, H) &= m^2 |H|^2 + \frac{1}{2}\lambda_H |H|^4 + A |H|^2\, S
\notag\\[5pt]
&\qquad
+ \frac12 (M^2 +\kappa |H|^2)\, S^2 + \frac{1}{3!}\mu_S S^3 + \frac{1}{4!} \lambda_S S^4 \,.
\end{align}

\subsection*{A New Field Basis}

In the Lagrangian of Eq.~\eqref{eq:L_singlet}, the heavy field $S$ couples to the Standard Model Higgs via nonderivative operators in the potential, while the metric is trivial. We can equally adopt a field basis where all interactions of the heavy field are encoded in a nontrivial metric. This is achieved via a nonderivative field redefinition for the heavy field:
\begin{align}
S = f_0\bigl(|H|^2\bigr) + f_1\bigl(|H|^2\bigr)\,\widetilde{S} + f_2(|H|^2)\,\widetilde{S}^2 + f_3\bigl(|H|^2\bigr)\,\widetilde{S}^3 + \cdots \,.
\label{eq:S_redef}
\end{align}
Requiring
\begin{align}
\widetilde{V}\bigl(\widetilde{S}, H\bigr) = V\bigl( S(\widetilde{S}, H) , H \bigr) = \widetilde{V}_0 (|H|^2) + \frac12 M^2\widetilde{S}^2 \,,
\label{eq:tV}
\end{align}
we can fix the coefficient functions $f_n(|H|^2)$ in \cref{eq:S_redef}, order by order in $1/M$ for each of them. Concretely, we match the Taylor expansion coefficients around $\tS=0$ on both sides of \cref{eq:tV}. At $\mathcal{O}(\tS)$, we have:
\begin{equation}
\frac{\partial\widetilde{V}}{\partial\tS} \biggr|_{\tS=0} = \frac{\partial V}{\partial S} \biggr|_{S=f_0(|H|^2)} f_1(|H|^2) = 0 \,.
\end{equation}
Requiring $f_1(|H|^2)\ne 0$, we see that $f_0(|H|^2)$ solves the heavy field's EOM at $\mathcal{O}(\partial^0)$, and we obtain:
\begin{align}
f_0\bigl(|H|^2\bigr) &= - \frac{A}{M^2}|H|^2 + \frac{A}{M^4}\biggl(\kappa - \frac{\mu_S A}{2M^2}\biggr)|H|^4
\notag\\[5pt] 
&\quad
- \frac{A}{M^6}\Biggl[\biggl(\kappa - \frac{\mu_S A}{M^2}\biggr)\biggl(\kappa - \frac{\mu_S A}{2M^2}\biggr) - \frac{\lambda_S A^2}{6M^2}\Biggr]|H|^6 + \mathcal{O}\bigl(|H|^8\bigr) \,.
\label{eq:singlet/result_f0}
\end{align}
The remaining coefficient functions $f_n(|H|^2)$ ($n\ge 1$) can be determined in the same way:
\begin{subequations}
\begin{align}
\frac{\partial^2\widetilde{V}}{\partial\tS^2} \biggr|_{\tS = 0} = M^2
&\quad\Longrightarrow\quad
f_1(|H|^2) = M\, \biggl(\frac{\pd^2 V}{\pd S^2}\biggr)^{-\frac12} \biggr|_{S = f_0} \,, \\[8pt]
\frac{\partial^3\widetilde{V}}{\partial\tS^3}\biggr|_{\tS = 0} = 0
&\quad\Longrightarrow\quad
f_2(|H|^2) = - \frac16 \left(\frac{\pd^3 V}{\pd S^3}\right) \left(\frac{\pd^2 V}{\pd S^2}\right)^{-1} \biggr|_{S = f_0}f_1^2 \,,
\end{align}
\end{subequations}
and so on. Note that the minimum operator dimension of $f_n(|H|^2)$ is 2 for $n=0$ (see \cref{eq:singlet/result_f0}) and 0 for all $n\ge1$:
\begin{subequations}
\begin{align}
f_1(|H|^2) &=  1 + \cdots \,, \\[5pt]
f_2(|H|^2) &= -\frac{\mu_S}{6M^2} + \cdots \,, \text{ etc.}
\end{align}
\end{subequations}

Further writing the Standard Model Higgs field in terms of its real components in the exponential parameterization,
\begin{align}\label{eq:singlet/spherical}
H = \frac{1}{\sqrt{2}}\,\tH\, e^{i\pi^a\sigma^a/v}
\left(\begin{array}{c}
0 \\
1
\end{array}\right) \,,
\end{align}
we obtain the UV Lagrangian in the $(\widetilde{S}, \tH, \pi^a)$ basis:
\begin{equation}
\mathcal{L} = \frac{1}{2} \,
\bigl(\partial_\mu\widetilde{S} \;\; \partial_\mu\tH \;\; \partial_\mu\pi^a \bigr)
\begin{pmatrix}
g_{SS} & g_{Sh} & 0 \\
g_{hS} & g_{hh} & 0 \\
0 & 0 & g_{ab}
\end{pmatrix}
\begin{pmatrix}
\partial^\mu\widetilde{S} \\ \partial^\mu\tH \\ \partial^\mu\pi^b
\end{pmatrix}
-\frac{1}{2}\, M^2\,\widetilde{S}^2 -W(\tH) \,,
\end{equation}
where
\begin{subequations}\label{eq:singlet/g}
\begin{align}
g_{SS} &= f_1^2 + 4f_1f_2\widetilde{S} + O(\widetilde{S}^2) \,, \\[8pt]
g_{Sh} = g_{hS} &= \left[f_1f_0^{(1)} + (2f_2f_0^{(1)} + f_1f_1^{(1)})\widetilde{S} + O(\widetilde{S}^2)\right] \tH \,, \\[8pt]
g_{hh} &= 1 + \left[(f_0^{(1)})^2 + 2f_0^{(1)} f_1^{(1)} \widetilde{S} + O(\widetilde{S}^2)\right]\tH^2 \,, \\[8pt]
g_{ab} &= \frac{\tH^2}{v^2}\,\hat{g}_{ab} \qquad \text{with} \;\; \hat{g}_{ab} \;\;\text{given by Eq.~\eqref{eq:sigma/metric_on_sphere}} \,, \\[8pt]
W(\tH) &= \widetilde{V}_0(\tH^2/2) = V(S = f_0, H) \,.
\end{align}
\end{subequations}
Here and in what follows, $f_n^{(m)}$ denotes the $m$-th derivative of $f_n(|H|^2)$ with respect to $|H|^2$, and the substitution $|H|^2 = \tH^2/2$ is implicit. We use $S$, $h$ and $a,b,\cdots$ for indices associated with the field manifold coordinates $\widetilde{S}$, $\tH$ and $\pi^a, \pi^b, \cdots$.

What this field redefinition buys us is clear from the discussion in Sec.~\ref{sec:guolea} on possible simplifying assumptions about the UV theory. In the new basis $(\widetilde{S}, \tH, \pi^a)$, nonderivative interactions of the heavy field $\widetilde{S}$ have been eliminated in favor of derivative interactions (as encoded in a nontrivial metric). Meanwhile, the singlet extended Standard Model enjoys a $\mathbb{Z}_2$ symmetry under which $\tH\to-\tH$, $\pi^a\to -\pi^a$. So the operator dimension counting for the $\,\U$ matrix follows Eq.~\eqref{eq:mindim_z2_noPhi}, and only the \ca{highlighted} terms in Tables~\ref{tab:guolea/0d_ab}-\ref{tab:guolea/4d_ab} are needed for one-loop matching up to dimension 6.

\subsection*{Tree-Level Matching}

For tree-level matching, since $[\widetilde{S}_\c] = 4$, only operators of dimension-8 or higher are generated by replacing $\widetilde{S}=\widetilde{S}_\c$, so we have
\begin{align}
\mathcal{L}_\text{EFT}^{[0]} &= \frac12\, \bigl(\partial_\mu\tH \;\; \partial_\mu\pi^a \bigr)
\begin{pmatrix}
g_{hh} & 0 \\
0 & g_{ab}
\end{pmatrix}
\begin{pmatrix}
\partial^\mu\tH \\ \partial^\mu\pi^b
\end{pmatrix}
-W(\tH) + (\text{dim-8})
\notag\\[8pt]
&= \frac12\, \bigl(\partial_\mu\tH \;\; \partial_\mu\pi^a \bigr)
\begin{pmatrix}
1 & 0 \\
0 & g_{ab}
\end{pmatrix}
\begin{pmatrix}
\partial^\mu\tH \\ \partial^\mu\pi^b
\end{pmatrix}
+\frac{A^2}{2M^4} \,\tH^2 (\partial_\mu\tH)^2
\notag\\[5pt]
&\quad
- \frac{m^2}{2} \,\tH^2 -\frac{1}{8}\biggl(\lambda_H-\frac{A^2}{M^2}\biggr) \,\tH^4 -\frac{A^2}{16M^4}\biggl(\kappa - \frac{\mu_S A}{3M^2} \biggr) \,\tH^6 + (\text{dim-8}) \,.
\end{align}
Writing the result in terms of the original Higgs doublet field $H$, we have
\begin{align}
\mathcal{L}_\text{EFT}^{[0]} &= |\partial_\mu H|^2 -m^2 |H|^2 -\frac{1}{2}\biggl(\lambda_H-\frac{A^2}{M^2}\biggr) |H|^4 \notag\\[5pt]
&\quad -\frac{A^2}{2M^4}\biggl(\kappa - \frac{\mu_S A}{3M^2} \biggr)\, |H|^6 + \frac{A^2}{2M^4} \bigl(\partial_\mu|H|^2\bigr)^2 + (\text{dim-8}) \,,
\end{align}
in agreement with previous calculations in the literature, \eg\ Refs.~\cite{deBlas:2017xtg,Jiang:2018pbd,Haisch:2020ahr,Cohen:2020fcu,Li:2023cwy}.

\subsection*{One-Loop Matching}

To obtain the EFT Lagrangian at one-loop level, we compute the Christoffel symbols from the metric whose nonzero components are given above in \cref{eq:singlet/g}; see App.~\ref{app:singlet} for details. We then derive the $\U$, $\Pd_\mu$, $\K_\mu$ and $\Y_{\mu\nu}$ matrices which enter the UOLEA. The results for the nonzero entries of these matrices, up to the order needed to derive EFT operators up to dimension-6, are
\begin{subequations}
\begin{align}
\U^S{}_S = \, & \, M^2 \frac{1}{f_1^2} \biggl\{1 - f_1^2 + \bigl(f_0^{(1)}\bigr)^2\tH^2 - \frac{3}{f_1}\Bigl[2f_2 + f_0^{(1)}\bigl(2f_2f_0^{(1)} - f_1f_1^{(1)}\bigr) \,\tH^2\Bigr]\tS_\c\biggr\} \nonumber \\[5pt]
& + (\text{dim-7}), \\[5pt]
\U^S{}_h = \, & \, -\frac{f_0^{(1)}}{f_1} \,\tH\,\bigl(  \widetilde{V}_0^{(2)}\tH^2 + \widetilde{V}_0^{(1)} \bigr)
- M^2\Biggl[\frac{f_1^{(1)}}{f_1^3} - \frac{\bigl(f_0^{(1)}\bigr)^2}{f_1^2}\Biggr]\tS_\c \tH + (\text{dim-6}) , \\[5pt]
\U^h{}_S = \, & \, -M^2\frac{f_0^{(1)}}{f_1}\tH + (\text{dim-4}), \qquad\quad
\U^h{}_h = \,  \widetilde{V}_0^{(2)}\tH^2 + \widetilde{V}_0^{(1)}
+ (\text{dim-3}) , \\[5pt]
\U^a{}_b = \, & \,  \widetilde{V}_0^{(1)}\, \delta^a_b, \\[5pt]
(\Pd_\mu)^S{}_S = \, & \, i\,\pd_\mu + i\,\frac{f_1^{(1)}}{f_1}\tH\bigl(\pd_\mu\tH\bigr) + (\text{dim-4}), \\[5pt]
(\Pd_\mu)^h{}_h = \, & \, i\,\pd_\mu + (\text{dim-4}), \qquad \qquad
(\Pd_\mu)^h{}_b = -i\,\frac{\tH}{v^2}\,\hat{g}_{bc}(\pd_\mu\pi^c) + (\text{dim-8}), \\[5pt]
(\Pd_\mu)^a{}_h = \, & \, i\,\frac{1}{\tH}(\pd_\mu\pi^a), \qquad\qquad
(\Pd_\mu)^a{}_b = i\,\big(\hat{\scr D}_\mu\bigr)^a{}_b + i\,\frac{1}{\tH}\,\delta^a_b\bigl(\pd_\mu\tH\bigr), \\[5pt] 
(\K_\mu)^S{}_h = \, & \, i\,\frac{1}{f_1}\bigl(f_0^{(1)} + f_0^{(2)}\tH^2\bigr)\bigl(\pd_\mu\tH\bigr) + (\text{dim-5}), 
\\[5pt]
(\K_\mu)^S{}_b = \, & \, i\,\frac{f_0^{(1)}}{f_1}\frac{\tH^2}{v^2}\,\hat{g}_{bc} \,(\pd_\mu\pi^c) + (\text{dim-5}), 
\end{align}
\end{subequations}
where
\begin{align}
\tS_\c = -\frac{1}{M^2}\biggl(1 - \frac{\pd^2}{M^2}\biggr)\biggl[f_1f_0^{(1)}\Bigl(\tH\bigl(\pd^2\tH\bigr) + \bigl(\pd_\mu\tH\bigr)^2\Bigr) + f_1f_0^{(2)}\tH^2\bigl(\pd_\mu\tH\bigr)^2\biggr] + (\text{dim-8}) \,,
\end{align}
$\hat{\scr D}_\mu$ is defined with respect to the connection $\hat{\Gamma}^a_{bc}$ in \cref{eq:sigma_connection_abc}, and $\tV_0^{(m)}$ denotes the $m$-th derivative of $\tV_0$ with respect to $|H|^2$ (the substitution $|H|^2 = \tH^2/2$ is implicit). Other entries of the $\,\U$, $\Pd_\mu$, $\K_\mu$ matrices, as well as all entries of the $\,\Y_{\mu\nu}$ matrix, either vanish or start at sufficiently high operator dimensions that they do not contribute to EFT matching up to dimension 6. Note also that despite a nontrivial metric and nonzero connection in the $(\widetilde{S}, \tH, \pi^a)$ basis, the Riemann curvature tensor (which enters $\,\U$ and $\,\Y_{\mu\nu}$) vanishes as it must since the field manifold is flat (as is clear from the original Lagrangian Eq.~\eqref{eq:L_singlet}).

Substituting these expressions into the \ca{highlighted} entries in Tables~\ref{tab:guolea/0d_ab}-\ref{tab:guolea/4d_ab}, we obtain the contribution to $\mathcal{L}_\text{EFT}^{[1]}$ from each term in the UOLEA (truncated at dimension 6):
\begin{subequations}
\label{eq:singlet/result}
\begin{align}
\op{tr} \bigl(\,\U_{HH}\bigr) =&\;\;
\frac{1}{2}\,\biggl( \kappa - \frac{\mu_S A}{M^2} + \frac{2A^2}{M^2} \biggr) \,\tH^2 \nonumber \\[5pt]
& + \frac{A}{8M^3}\biggl[ \frac{2\kappa\mu_S}{M} + \frac{4\mu_S A^2}{M^3} - \frac{A}{M} \biggl( 12 \kappa + \frac{\mu_S^2}{M^2} - \lambda_S \biggr) \biggr] \,\tH^4 \nonumber \\[5pt]
& - \frac{A}{48M^5} \biggl[ \frac{6\kappa^2\mu_S}{M} + \frac{6A^3}{M^3}\biggl(\lambda_S - \frac{3\mu_S^2}{M^2}\biggr) 
+ \frac{\mu_S A^2}{M^3}\biggl(72\kappa - 4\lambda_S + \frac{3\mu_S^2}{M^2}\biggr) \nonumber \\[5pt]
& \qquad\qquad - \frac{3\kappa A}{M}\biggl(24\kappa - 2\lambda_S + \frac{3\mu_S^2}{M^2}\biggr) \biggr] \,\tH^6 \nonumber \\[5pt]
& + \frac{A}{M^5}\biggr[\frac{\mu_S A^2}{M^3} + \frac{\kappa\mu_S}{M} - \frac{A}{M}\biggl(3\kappa - \lambda_S + \frac{\mu_S^2}{M^2}\biggr)\biggr] \,\tH^2\bigl(\pd_\mu\tH\bigr)^2 \,, \\[10pt]
\op{tr} \bigl((\,\U_{HH})^2\bigr) =&\;\;
\frac{1}{4}\biggl(\kappa - \frac{\mu_S A}{M^2} + \frac{2A^2}{M^2}\biggr)^2 \,\tH^4 \nonumber \\[5pt]
& + \frac{A}{8M^3} \biggl(\kappa - \frac{\mu_S A}{M^2} + \frac{2A^2}{M^2}\biggr) \biggl[ \frac{2\kappa\mu_S}{M} + \frac{4\mu_S A^2}{M^3} \nonumber \\[5pt]
& \qquad\qquad - \frac{A}{M} \biggl( 12 \kappa + \frac{\mu_S^2}{M^2} - \lambda_S \biggr) \biggr] \,\tH^6 \nonumber \\[5pt]
& - \frac{2\,\mu_SA}{M^4} \biggl(\kappa - \frac{\mu_S A}{M^2} + \frac{2A^2}{M^2}\biggr)\, \tH^2\bigl(\pd_\mu\tH\bigr)^2 \,, \\[10pt]
\op{tr} \bigl(\,\U_{HL}\,\U_{LH}\bigr) = 
\,&\, \frac{A^2}{M^2}m^2 \tH^2 + \frac{3A^2}{2M^2}\biggl[ \lambda_H - \frac{A^2}{M^2} - \frac{m^2}{M^2}\biggl(\kappa  - \frac{\mu_SA}{3M^2}\biggr)\biggr] \, \tH^4 \nonumber \\[5pt]
& - \frac{3A^2}{8M^4}\biggl(6\lambda_H - \frac{11A^2}{M^2}\biggr) \biggl(\kappa - \frac{\mu_SA}{3M^2}\biggr) \, \tH^6 \nonumber \\[5pt]
& - \frac{A^2}{M^4} \biggl(\kappa - \frac{\mu_S A}{M^2} + \frac{2A^2}{M^2}\biggr) \, \tH^2\bigl(\pd_\mu\tH\bigr)^2 \,, \\[10pt]
\op{tr} \bigl((\,\U_{HH})^3\bigr)
= \,&\, \frac{1}{8}\biggl(\kappa - \frac{\mu_S A}{M^2} + \frac{2A^2}{M^2}\biggr)^3 \, \tH^6 \,, \\[10pt]
\op{tr} \bigl(\,\U_{HH} \,\U_{HL} \,\U_{LH}\bigr)
= \,&\, \frac{A^2}{4M^2}\biggl(\kappa - \frac{\mu_S A}{M^2} + \frac{2A^2}{M^2}\biggr) \biggl[2m^2 + 3\biggl(\lambda_H - \frac{A^2}{M^2}\biggr) \,\tH^2\biggr] \, \tH^4 \,, \\[10pt]
\op{tr} \bigl(\,\U_{HL} \,\U_{LL} \,\U_{LH}\bigr)
= \,&\, \frac{A^2}{4M^2}\biggl[2m^2 + 3\biggl(\lambda_H - \frac{A^2}{M^2}\biggr) \,\tH^2\biggr]^2 \, \tH^2 \,, 
\end{align}
\begin{align}
\op{tr} \bigl([\Pd_\mu, \U]_{HH} [\Pd^\mu, \,\U]_{HH}\bigr) = 
\,&\, -\biggl(\kappa - \frac{\mu_S A}{M^2} + \frac{2A^2}{M^2}\biggr)^2 \, \tH^2\bigl(\pd_\mu\tH\bigr)^2 \,, \\[10pt]
\op{tr} \bigl([\Pd_\mu, \U]_{HL} [\Pd^\mu, \U]_{LH}\bigr)
= \,& -\frac{A^2}{M^2}\biggl[ m^2 + \frac{9}{2} \biggl(\lambda_H - \frac{A^2}{M^2}\biggr) \, \tH^2 \biggr]\bigl(\pd_\mu\tH\bigr)^2 \nonumber \\[5pt]
& - \frac{A^2}{M^2}\biggl[ m^2 + \frac{3}{2}\biggl( \lambda_H - \frac{A^2}{M^2}\biggr) \, \tH^2 \biggr] \,\frac{\tH^2}{v^2}\,\hat{g} \,, \\[10pt]
\op{tr} \bigl((\K_{\mu})_{HL} \,\U_{LH} \,(\K^\mu)_{HL} \,\U_{LH}\bigr)
= \,& - \frac{A^4}{M^4} \,\tH^2\bigl(\pd_\mu\tH\bigr)^2 \,, \\[10pt]
\op{tr} \bigl([\Pd_\mu, \K^\mu]_{HL} \,U_{LH}\bigr)
= \,& -\frac{A^2}{M^2} \biggl[\bigl(\pd_\mu\tH\bigr)^2 + \frac{\tH^2}{v^2} \,\hat{g}\biggr] \nonumber \\[5pt]
& + \frac{3A^2}{2M^4}\biggl(\kappa - \frac{\mu_SA}{3M^2}\biggr) \, \tH^2 \biggl[3\bigl(\pd_\mu\tH\bigr)^2 + \frac{\tH^2}{v^2} \,\hat{g}\biggr] \,, \\[10pt]
\op{tr} \bigl([\Pd_\mu, \U]_{HH} \,(\K^\mu)_{HL} \,\U_{LH}\bigr)
= \,& \frac{A^2}{M^2}\biggl(\kappa - \frac{\mu_S A}{M^2} + \frac{2A^2}{M^2}\biggr) \, \tH^2\bigl(\pd_\mu\tH\bigr)^2 \,, \\[10pt]
\op{tr} \bigl((\K^\mu)_{HL} [\Pd_\mu, \U]_{LL} \,\U_{LH}\bigr)
= \,&\, \frac{A^2}{M^2}\biggl(\lambda_H - \frac{A^2}{M^2}\biggr) \, \tH^2 \biggl[3\bigl(\pd_\mu\tH\bigr)^2 + \frac{h^2}{v^2} \,\hat{g}\biggr] \,, \\[10pt]
\op{tr} \bigl((\K^\mu)_{HL} \,\U_{LL} [\Pd_\mu, \U]_{LH}\bigr)
= \,&\, \frac{A^2}{M^2} \biggl[ m^2 + \frac{3}{2}\biggl(\lambda_H - \frac{A^2}{M^2}\biggr) \,\tH^2 \biggr] \bigl(\pd_\mu\tH\bigr)^2  \nonumber \\[5pt]
& + \frac{A^2}{M^2} \biggl[ m^2 + \frac{1}{2}\biggl(\lambda_H - \frac{A^2}{M^2}\biggr) \,\tH^2 \biggr] \,\frac{\tH^2}{v^2}\,\hat{g} \, \\[10pt]
\op{tr} \bigl([\Pd_\mu, \K^\mu]\big._{HL} \bigl[\Pd_\nu, [\Pd^\nu ,\U]\bigr]_{LH}\bigr) = 
\,& -\frac{A^2}{M^2} \biggl\{ \bigl(\pd^2\tH\bigr)^2 - \frac{2}{v^2}\,\tH\bigl(\pd^2\tH\bigr)\,\hat{g} + \frac{4}{v^2}\bigl(\pd^\mu\tH\bigr)\bigl(\pd^\nu\tH\bigr)\,\hat{g}_{\mu\nu} \nonumber \\[5pt]
& + \frac{\tH^2}{v^2}\biggl[ \frac{1}{v^2}\,\hat{g}^2 - \bigl(\pd^\mu\pd^\nu\hat{g}_{\mu\nu}\bigr) + \hat{g}_{ab}\bigl(\hat{\scr D}_\mu\pd_\nu\pi\bigr)^a\bigl(\hat{\scr D}^\mu\pd^\nu\pi\bigr)^b \nonumber \\[5pt]
& - \hat{R}_{abcd}(\pd_\mu\pi^a)(\pd_\nu\pi^b)(\pd^\mu\pi^c)(\pd^\nu\pi^d) \biggr] \biggr\} \,, \label{eq:d4h2}
\end{align}
\end{subequations}
where $\hat{g}$, $\hat{g}_{\mu\nu}$ were defined in Eq.~\eqref{eq:ghat}. In the equations above, we have evaluated the traces in the $(\tS, \tH, \pi^a)$ basis. Using:
\begin{align}
\tH^2 = 2\, |H|^2 \,,\qquad
\tH^2\, \bigl(\pd_\mu \tH\bigr)^2 = \bigl(\pd_\mu |H|^2\bigr)^2 \,,\qquad
\bigl(\pd_\mu \tH\bigr)^2 + \frac{\tH^2}{v^2}\,\hat{g} = 2\, |\pd_\mu H|^2 \,,
\end{align}
and identifying the expression in curly brackets in Eq.~\eqref{eq:d4h2} as $2\,|\pd^2 H|^2$, we can rewrite the operators into SMEFT form. Adding up all the results from \cref{eq:singlet/result} multiplied by the respective coefficients listed in Tables~\ref{tab:guolea/0d_ab}-\ref{tab:guolea/4d_ab}, we find the one-loop-level EFT Lagrangian $\mathcal{L}_\text{EFT}^{[1]}$, which is in full agreement with results in the previous literature~\cite{Jiang:2018pbd,Haisch:2020ahr,Cohen:2020fcu}. This is expected since the field redefinition we performed to go to the $(\tS, \tH, \pi^a)$ basis, \cref{eq:S_redef,eq:tV}, does not mix the heavy field into the light fields.

It is interesting to note that while in the original field basis, where interactions of the heavy singlet field are all contained in the potential and the metric is trivial, one needs to evaluate 26 terms in the UOLEA to obtain the matching result up to dimension-6 (when gauge and Yukawa couplings are turned off). In contrast, only 14 terms in the geometric UOLEA contribute in the new basis where nonderivative interactions of the heavy singlet are eliminated at the expense of a nontrivial metric. Essentially, we have traded relevant and marginal operators without derivatives for irrelevant operators with derivatives in the UV theory, thereby reducing the number of possibilities when enumerating functional traces at the desired EFT operator dimension.

\section{Conclusion}
\label{sec:summary}

We have presented a procedure for integrating out heavy degrees of freedom in scalar field theories in the framework of field space geometry, and derived universal formulae for the EFT action up to one-loop level. The universality of one-loop EFT matching results following from functional methods has led to the Universal One-Loop Effective Action (UOLEA) program in recent years. The goal of this program is to perform the universal steps of functional matching calculations once and for all under general parameterizations of the UV Lagrangian, thereby eliminating the need for repetitive evaluations of functional determinants for specific UV theories. Our results here further extend the UOLEA by enlarging the space of UV theories to include two-derivative interactions.

For the special case of sigma models, the UOLEA up to 4-derivative terms is given by \cref{eq:sigma/loop_result} (together with tree-level matching result up to 6-derivative terms given by \cref{eq:sigma/tree_result}), while for general two-derivative scalar field theories, the UOLEA terms up to dimension-6 are shown in Tables~\ref{tab:guolea/0d_ab}-\ref{tab:guolea/0d_c}. These results are written in terms of quantities that transform covariantly under nonderivative redefinitions of the light fields, which correspond to coordinate transformations on the EFT submanifold.

The packaging of UOLEA operators into geometric objects brings both technical simplifications and potentially new insights into structures of EFTs. In the linear sigma model, for example, calculating just a handful of terms gives the entire tower of 2- and 4-derivative EFT operators, up to arbitrarily high canonical dimensions. In the singlet extended Standard Model, we saw in Sec.~\ref{sec:singlet} that working in a field basis where relevant and marginal nonderivative interactions of the heavy field are traded for irrelevant derivative interactions as encoded in a nontrivial metric reduces the UOLEA terms that need to be evaluated from 26 to 14. The situation here is reminiscent of the symmetry analysis of a recently discovered magic zero~\cite{Craig:2021ksw} -- the surprising cancellation in a matching coefficient in a vector-like fermions model~\cite{Arkani-Hamed:2021xlp} -- where a similar field redefinition was utilized to reduce the number of possible spurion combinations contributing to the given EFT operator. We are therefore hopeful that the geometric matching framework developed in this work can also be useful for systematically identifying additional magic zeroes~\cite{DelleRose:2022ygn,Bao:2024zzc} and shedding light on hidden symmetry constraints on the expected size of EFT matching coefficients. This synergies with recent efforts to calculate anomalous dimensions in the geometric framework~\cite{Helset:2022pde,Assi:2023zid,Jenkins:2023rtg,Jenkins:2023bls} which seek to further understand selection rules in EFT operator mixing~\cite{Alonso:2014rga, Elias-Miro:2014eia, Cheung:2015aba, Bern:2019wie, Craig:2019wmo, Jiang:2020rwz, EliasMiro:2020tdv, Baratella:2020lzz, Bern:2020ikv, Cao:2021cdt, Chala:2023jyx, Cao:2023adc, Chala:2023xjy}.

We have focused on scalar theories in this work as a first step to setting up the geometric matching formalism. It would be interesting to extend the geometric UOLEA to include higher-spin fields following recent works that extended field space geometry to incorporate fermion and vector fields~\cite{Finn:2019aip, Finn:2020nvn, Pilaftsis:2022las, Helset:2022tlf, Gattus:2023gep, Assi:2023zid}. Another future direction is to develop a covariant matching formalism on the functional manifold, where coordinate transformations include the most general field redefinitions involving derivatives~\cite{Cohen:2022uuw, Cheung:2022vnd, Cohen:2023ekv, Cohen:2024bml}.
In the singlet extended Standard Model discussed in Sec.~\ref{sec:singlet}, the field redefinition we considered does not mix the heavy field into the light fields, so the matching results before and after the field redefinition are in the same EFT operator basis. But more general field redefinitions in the UV theory would result in different EFT operator bases that are related by derivative field redefinitions; see \eg\ Ref.~\cite{Cohen:2020xca}. It would be especially interesting to systematically understand the relation between UV and EFT operator bases in a geometric framework that accommodates derivative field redefinitions.

\acknowledgments
\addcontentsline{toc}{section}{\protect\numberline{}Acknowledgments}
We thank Tim Cohen for helpful discussions and feedback on a preliminary draft.
X.-X.L.\ and Z.Z.\ are supported in part by the U.S.~National Science Foundation under grant PHY-2412880.
X.L.\ is supported in part by the U.S.~Department of Energy Grant No.DE-SC0009919 and in part by Simons Foundation Award 568420. 

\appendix
\section{Deriving the UOLEA for $O(N)$ Sigma Models}
\label{app:sigma}

In this appendix, we present additional details in the derivation of the UOLEA for sigma models in Sec.~\ref{sec:sigma}. As discussed in the main text, log-type traces do not contribute since $(\Y_{\mu\nu})^h{}_h=0$. For the power-type traces, we label each term in the series in the last two lines of Eq.~\eqref{eq:geo/eft_lag_expand} by the numbers of $\,\widetilde{\U}$, $\bigl\{ q_\mu - \Yt_{\mu\nu}\partial_q^\nu, \widetilde \K^\mu \bigr\}$ and $-\widetilde \K_\mu \widetilde \K^\mu$ factors, denoted by $(r,s,t)$. For example, the term
\begin{align}
-\frac{i}{2}\frac{1}{2}\int\frac{\md^dq}{(2\pi)^d} \op{tr} \left[ \tfrac{1}{q^2 - {\bf M}^2} \,\widetilde{\U} \tfrac{1}{q^2 - {\bf M}^2} \left\{q^\mu, \Yt_{\mu\nu}\pd_q^\nu\right\} \tfrac{1}{q^2 - {\bf M}^2} \bigl(-\widetilde{\K}_\mu\widetilde{\K}^\mu\bigr) \right]
\end{align}
is labeled as $(r,s,t) = (1, 0, 1)$. Since we aim to derive the UOLEA up to ${\cal O}(\pd^4)$, we can set a cutoff for $(r,s,t)$. From the derivative orders of $\,\U$, $\K$ and $\Y$ shown in Eq.~\eqref{eq:sigma/PowerCountingOfGeometricQuantities} (which are the minimum derivative orders of $\,\widetilde{\U}$, $\widetilde{\K}$, $\widetilde{\Y}$) we see that we need $2r + s + 2t \leq 4$. Also, since $\K$ has only off-diagonal blocks while all other matrices are block-diagonal, $s$ must be even for the trace to be nonzero. We therefore end up with 9 possible combinations of $(r,s,t)$: $(0,0,1)$, $(0,0,2)$, $(0,2,0)$, $(0,2,1)$, $(0,4,0)$, $(1,0,0)$, $(1,0,1)$, $(1,2,0)$, $(2,0,0)$. The evaluation of momentum integrals follows standard methods (see \eg\ Ref.~\cite{Cohen:2019btp}). We find for each set of traces (truncated at ${\cal O}(\pd^4)$):
\begin{subequations}
\begin{align}
{\cal L}_{\text{eff}}^{[1],(0,0,1)} 
= \,& -\frac{i}{2}\int\frac{\md^dq}{(2\pi)^d} \op{tr} \biggl[ \tfrac{1}{q^2 - {\bf M}^2} \bigl(-\widetilde{\K}_\mu\widetilde{\K}^\mu\bigr) + \tfrac{1}{q^2-{\bf M}^2} \bigl\{ q^\mu , \Yt_{\mu\nu} \partial_q^\nu \bigr\} \tfrac{1}{q^2 - {\bf M}^2} \bigl(-\widetilde{\K}_\mu\widetilde{\K}^\mu\bigr) \biggr] \nonumber \\[8pt]
= &~ \intmdx \frac{1}{16\pi^2} \biggl\{ -\tfrac{1-L}{2} M^2\bF_{,h}^2\,\hat{g} - (1-L)\bF_{,h}^2\bF_{,hh}\,\hat{g}^2 \biggr\}
\,, \\[15pt]
{\cal L}_{\text{eff}}^{[1],(0,0,2)} 
= \,& -\frac{i}{2}\frac{1}{2}\int\frac{\md^dq}{(2\pi)^d} \op{tr} \biggl[ \tfrac{1}{q^2 - {\bf M}^2} \bigl(-\widetilde{\K}_\mu\widetilde{\K}^\mu\bigr) \tfrac{1}{q^2 - {\bf M}^2} \bigl(-\widetilde{\K}_\mu\widetilde{\K}^\mu\bigr) \biggr]
\nonumber\\[8pt]
= \,&\, \intmdx \frac{1}{16\pi^2} \biggl\{ -\tfrac{L}{4} \bF_{,h}^4\,\hat{g}^2 \biggr\}
\,, \\[15pt]
{\cal L}_{\text{eff}}^{[1],(0,2,0)} 
= \,& -\frac{i}{2}\frac{1}{2}\int\frac{\md^dq}{(2\pi)^d} \op{tr} \biggl[ \tfrac{1}{q^2 - {\bf M}^2} \bigl\{ q_\mu - \Yt_{\mu\nu}\partial_q^\nu, \widetilde{\K}^\mu \bigr\} \tfrac{1}{q^2 - {\bf M}^2} \bigl\{ q_\mu - \Yt_{\mu\nu}\partial_q^\nu, \widetilde{\K}^\mu \bigr\} \nonumber \\[5pt]
& + \tfrac{1}{q^2-{\bf M}^2} \bigl\{ q^\mu , \Yt_{\mu\nu} \partial_q^\nu \bigr\} \tfrac{1}{q^2 - {\bf M}^2} \bigl\{ q_\mu - \Yt_{\mu\nu}\partial_q^\nu, \widetilde{\K}^\mu \bigr\} \tfrac{1}{q^2 - {\bf M}^2} \bigl\{ q_\mu - \Yt_{\mu\nu}\partial_q^\nu, \widetilde{\K}^\mu \bigr\} \nonumber \\[5pt]
& + \tfrac{1}{q^2 - {\bf M}^2} \bigl\{ q_\mu - \Yt_{\mu\nu}\partial_q^\nu, \widetilde{\K}^\mu \bigr\} \tfrac{1}{q^2-{\bf M}^2} \bigl\{ q^\mu , \Yt_{\mu\nu} \partial_q^\nu \bigr\} \tfrac{1}{q^2 - {\bf M}^2} \bigl\{ q_\mu - \Yt_{\mu\nu}\partial_q^\nu, \widetilde{\K}^\mu \bigr\} \biggr] \nonumber \\[5pt]
= \,&\, \intmdx \frac{1}{16\pi^2} \biggl\{\tfrac{3-2L}{4}M^2\bF_{,h}^2\,\hat{g} + \tfrac{3-2L}{2}\bF_{,h}^2\bF_{,hh}\,\hat{g}^2 \nonumber \\[5pt]
& \hspace{30pt} + \tfrac{4-3L}{18} \bF_{,h}^2\,\hat{g}_{ab}(\hat{\scr D}_\mu \pd^\mu \pi)^a(\hat{\scr D}_\nu \pd^\nu \pi)^b - \tfrac{5-6L}{36} \bF_{,h}^2\,\hat{g}_{ab}(\hat{\scr D}_\mu \pd_\nu \pi)^a(\hat{\scr D}^\mu \pd^\nu \pi)^b \nonumber \\[5pt]
& \hspace{30pt} + \tfrac{11-6L}{18} \bF_{,h}^2\hat{R}_{abcd}(\pd_\mu \pi^a)(\pd_\nu \pi^b)(\pd^\mu \pi^c)(\pd^\nu \pi^d) \biggr\}
\,, \\[15pt]
{\cal L}_{\text{eff}}^{[1],(0,2,1)} 
= \,& -\frac{i}{2}\frac{1}{3}\,3\int\frac{\md^dq}{(2\pi)^d} \op{tr} \biggl[ \tfrac{1}{q^2 - {\bf M}^2} \bigl\{ q_\mu - \Yt_{\mu\nu}\partial_q^\nu, \widetilde{\K}^\mu \bigr\} \nonumber \\[5pt]
& \qquad\qquad\qquad \times \tfrac{1}{q^2 - {\bf M}^2} \bigl\{ q_\mu - \Yt_{\mu\nu}\partial_q^\nu, \widetilde{\K}^\mu \bigr\} \tfrac{1}{q^2 - {\bf M}^2} \bigl(-\widetilde{\K}_\mu\widetilde{\K}^\mu\bigr) \biggr] \nonumber \\[5pt]
= & \intmdx \frac{1}{16\pi^2} \biggl\{ -\tfrac{1-2L}{4}\bF_{,h}^4\,\hat{g}^2 - \tfrac{3-2L}{4}\bF_{,h}^4\,\hat{g}_{\mu\nu}\hat{g}^{\mu\nu} \biggr\}
\,, \\[15pt]
{\cal L}_{\text{eff}}^{[1],(0,4,0)} 
= \,& -\frac{i}{2}\frac{1}{4}\int\frac{\md^dq}{(2\pi)^d} \op{tr} \biggl[ \tfrac{1}{q^2 - {\bf M}^2} \bigl\{ q_\mu - \Yt_{\mu\nu}\partial_q^\nu, \widetilde{\K}^\mu \bigr\} \tfrac{1}{q^2 - {\bf M}^2} \bigl\{ q_\mu - \Yt_{\mu\nu}\partial_q^\nu, \widetilde{\K}^\mu \bigr\} \nonumber \\[5pt]
& \qquad\qquad\qquad \times \tfrac{1}{q^2 - {\bf M}^2} \bigl\{ q_\mu - \Yt_{\mu\nu}\partial_q^\nu, \widetilde{\K}^\mu \bigr\} \tfrac{1}{q^2 - {\bf M}^2} \bigl\{ q_\mu - \Yt_{\mu\nu}\partial_q^\nu, \widetilde{\K}^\mu \bigr\} \biggr] \nonumber \\[5pt]
= \,&\, \intmdx \frac{1}{16\pi^2} \biggl\{ \tfrac{5-6L}{36}\bF_{,h}^4\,\hat{g}^2 + \tfrac{5-6L}{18} \bF_{,h}^4\hat{g}_{\mu\nu}\,\hat{g}^{\mu\nu} \biggr\}
\,, \\[15pt]
{\cal L}_{\text{eff}}^{[1],(1,0,0)} 
= \,& -\frac{i}{2}\int\frac{\md^dq}{(2\pi)^d} \op{tr} \biggl[ \tfrac{1}{q^2 - {\bf M}^2} \,\widetilde{\U} + \tfrac{1}{q^2-{\bf M}^2} \bigl\{ q^\mu , \Yt_{\mu\nu} \partial_q^\nu \bigr\} \tfrac{1}{q^2 - {\bf M}^2} \,\widetilde{\U} \biggr] \nonumber \\[5pt]
= \,&\, \intmdx \frac{1}{16\pi^2} \biggl\{ \tfrac{1-L}{2}(\bF_{,h}\bV_{,hhh} - M^2\bF_{,hh}) \,\hat{g} - \tfrac{1}{M^2} \Bigl(\tfrac{1-L}{2}\bF_{,h}\bV_{,hhh}\Bigr) (\pd^2\hat{g}) %
\nonumber \\[5pt]
& + \biggl[ -\tfrac{1-L}{2}\bF_{,h}(\bF_{,h}\bF_{,hh} + \bF_{,hhh}) + \tfrac{1}{M^2} \Bigl( \tfrac{1-L}{2}\bF_{,h}^3\bV_{,hhh} + \tfrac{1-L}{2}\bF_{,h}\bF_{,hh}\bV_{,hhh} \nonumber \\[5pt]
& \hspace{6em} + \tfrac{1-L}{4}\bF_{,h}^2\bV_{,hhhh}\Bigr) -\tfrac{1}{M^4} \Bigl(\tfrac{1-L}{4}\bF_{,h}^2\bV_{,hhh}^2\Bigr) \biggr] \,\hat{g}^2 \biggr\}
\,, \\[15pt]
{\cal L}_{\text{eff}}^{[1],(1,0,1)} = 
\,& -\frac{i}{2}\frac{1}{2}\,2\int\frac{\md^dq}{(2\pi)^d} \op{tr} \biggl[ \tfrac{1}{q^2 - {\bf M}^2} \,\widetilde{\U} \tfrac{1}{q^2 - {\bf M}^2} \bigl(-\widetilde{\K}_\mu\widetilde{\K}^\mu\bigr) \biggr] \nonumber \\[5pt]
= \,&\, \intmdx \frac{1}{16\pi^2} \biggl\{ \tfrac{L}{2M^2} \bF_{,h}^2(\bF_{,h}\bV_{,hhh} - M^2\bF_{,hh}) \,\hat{g}^2 \biggr\}
\,, \\[15pt]
{\cal L}_{\text{eff}}^{[1],(1,2,0)} = 
\,& -\frac{i}{2}\frac{1}{3}\,3\int\frac{\md^dq}{(2\pi)^d} \op{tr} \biggl[ \tfrac{1}{q^2 - {\bf M}^2} \,\widetilde{\U} \tfrac{1}{q^2 - {\bf M}^2} \bigl\{ q_\mu - \Yt_{\mu\nu}\partial_q^\nu, \widetilde{\K}^\mu \bigr\} \tfrac{1}{q^2 - {\bf M}^2} \bigl\{ q_\mu - \Yt_{\mu\nu}\partial_q^\nu, \widetilde{\K}^\mu \bigr\} \biggr] \nonumber \\[5pt]
= \,&\, \intmdx \frac{1}{16\pi^2} \biggl\{\tfrac{1-2L}{4M^2}\bF_{,h}^2(\bF_{,h}\bV_{,hhh} - M^2\bF_{,hh}) \,\hat{g}^2 + \tfrac{3-2L}{4}\bF_{,h}^4 \,\hat{g}_{\mu\nu}\hat{g}^{\mu\nu} \nonumber \\[5pt]
& \hspace{30pt} + \tfrac{3-2L}{4}\bF_{,h}^2\hat{R}_{abcd}(\pd_\mu \pi^a)(\pd_\nu \pi^b)(\pd^\mu \pi^c)(\pd^\nu \pi^d) \biggr\}
\,, \\[15pt]
{\cal L}_{\text{eff}}^{[1],(2,0,0)} = 
\,& -\frac{i}{2}\frac{1}{2}\int\frac{\md^dq}{(2\pi)^d} \op{tr} \biggl[ \tfrac{1}{q^2 - {\bf M}^2} \,\widetilde{\U} \tfrac{1}{q^2 - {\bf M}^2} \,\widetilde{\U} \biggr] \nonumber \\[5pt]
= \,& \intmdx \frac{1}{16\pi^2} \biggl\{ -\tfrac{L}{4M^4} (\bF_{,h}\bV_{,hhh} - M^2\bF_{,hh})^2 \,\hat{g}^2 \biggr\} \,.
\end{align}
\end{subequations}
Putting these results together gives us \cref{eq:sigma/loop_result} in the main text.

\section{Additional Details of the Singlet Extended Standard Model}
\label{app:singlet}

In the $\{\tS, \tH, \pi^a\}$ basis of the singlet extended Standard Model, the nonzero entries of the inverse metric are given by
\begin{subequations}
\begin{align}
g^{SS} &= \frac{1}{f_1^2}\bigl[1 + \bigl(f_0^{(1)}\bigr)^2\tH^2\bigr] - \frac{1}{f_1^3}\bigl[4f_2 + 2f_0^{(1)}\bigl(2f_2f_0^{(1)} - f_1f_1^{(1)}\bigr)\,\tH^2\bigr]\tS + O(\tS^2) \,, \\[5pt] 
g^{Sh} &= g^{hS} = -\frac{f_0^{(1)}}{f_1}\,\tH + \frac{1}{f_1^2}\bigl(2f_2f_0^{(1)} - f_1f_1^{(1)}\bigr) \,\tH\,\tS + O(\tS^2) \,, \\[5pt]
g^{hh} &= 1 + O(\tS^2) \,, \\[5pt]
g^{ab} &= \frac{v^2}{\tH^2}\, \hat{g}^{ab} \,.
\end{align}
\end{subequations}
The Christoffel symbols of the first kind are
\begin{subequations}
\begin{alignat}{2}
\Gamma_{SSS} &= 2f_1f_2 + O(\widetilde{S}) \,,\quad &
\Gamma_{Shh} &= f_1f_0^{(1)} + f_1f_0^{(2)}\tH^2 + O(\widetilde{S})
\,, \\[5pt]
\Gamma_{SSh} &= \Gamma_{ShS} = \bigl[f_1f_1^{(1)} + O(\widetilde{S})\bigr]\, \tH \,,\quad &&
\\[5pt]
\Gamma_{hSS} &= \bigl[2f_2f_0^{(1)} + O(\widetilde{S})\bigr] \,\tH \,,\quad &
\Gamma_{hhh} &= \biggl[\bigl(f_0^{(1)}\bigr)^2 
+ f_0^{(1)}f_0^{(2)}\tH^2 + O(\widetilde{S})\biggr]\, \tH
\,, \\[5pt]
\Gamma_{hSh} &= \Gamma_{hhS} = \bigl[f_0^{(1)}f_1^{(1)} + O(\widetilde{S})\bigr]\, \tH^2 \,,\quad &
\Gamma_{hab} &= -\frac{1}{v^2}\,\tH\,\hat{g}_{ab}
\,, \\[5pt]
\Gamma_{ahb} &= \Gamma_{abh} = \frac{1}{v^2}\,\tH\,\hat{g}_{ab} \,,\quad &
\Gamma_{abc} &= \frac{1}{v^2}\,\tH^2\,\hat{\Gamma}_{abc} \,,
\end{alignat}
\end{subequations}
and the Christoffel symbols of the second kind are
\begin{subequations}
\begin{alignat}{2}
\Gamma^S_{SS} &= 2\frac{f_2}{f_1} + O(\widetilde{S}) \,,\qquad &
\Gamma^S_{Sh} &= \Gamma^S_{hS} = \biggl[\frac{f_1^{(1)}}{f_1} + O(\widetilde{S})\biggr] \,\tH
\,, \\[5pt]
\Gamma^S_{hh} &= \frac{1}{f_1}\bigl(f_0^{(1)} + f_0^{(2)}\tH^2\bigr) + O(\widetilde{S}) \,,\qquad &
\Gamma^S_{ab} &= \biggl[\frac{f_0^{(1)}}{f_1} + O(\tS)\biggr]\frac{1}{v^2}\,\tH^2\,\hat{g}_{ab}
\,, \\[5pt]
\Gamma^h_{SS} &= O(\widetilde{S}^2) \,,\qquad &
\Gamma^h_{Sh} &= \Gamma^h_{hS} = O(\widetilde{S}^2)
\,, \\[5pt]
\Gamma^h_{hh} &= O(\widetilde{S}) \,,\qquad &
\Gamma^h_{ab} &= -\frac{1}{v^2}\tH\,\hat{g}_{ab} + O(\widetilde{S}^2)
\,, \\[5pt]
\Gamma^a_{hb} &= \frac{1}{\tH}\,\delta^a_b \,,\qquad &
\Gamma^a_{bc} &= \hat{\Gamma}^a_{bc} \,.
\end{alignat}
\end{subequations}


\bibliographystyle{JHEP}
\bibliography{ref}

\providecommand{\href}[2]{#2}\begingroup\raggedright\begin{thebibliography}{10}

\bibitem{Borchers1960}
H.-J. Borchers, {\it {{\"U}ber die Mannigfaltigkeit der interpolierenden Felder
  zu einer kausalen S-Matrix}},  {\em Il Nuovo Cimento (1955-1965)} {\bf 15}
  (1960), no.~5 784--794.

\bibitem{Chisholm:1961tha}
J.~S.~R. Chisholm, {\it {Change of variables in quantum field theories}},  {\em
  Nucl. Phys.} {\bf 26} (1961), no.~3 469--479.

\bibitem{Kamefuchi:1961sb}
S.~Kamefuchi, L.~O'Raifeartaigh, and A.~Salam, {\it {Change of variables and
  equivalence theorems in quantum field theories}},  {\em Nucl. Phys.} {\bf 28}
  (1961) 529--549.

\bibitem{tHooft:1973wag}
G.~'t~Hooft and M.~J.~G. Veltman, {\it {DIAGRAMMAR}},  {\em NATO Sci. Ser. B}
  {\bf 4} (1974) 177--322.

\bibitem{Arzt:1993gz}
C.~Arzt, {\it {Reduced effective Lagrangians}},  {\em Phys. Lett. B} {\bf 342}
  (1995) 189--195, [\href{http://arxiv.org/abs/hep-ph/9304230}{{\tt
  hep-ph/9304230}}].

\bibitem{Epstein2008}
H.~Epstein, {\it On the borchers class of a free field},  {\em Il Nuovo Cimento
  (1955-1965)} {\bf 27} (2008), no.~4 886.

\bibitem{Manohar:2018aog}
A.~V. Manohar, {\it {Introduction to Effective Field Theories}},
  \href{http://arxiv.org/abs/1804.05863}{{\tt arXiv:1804.05863}}.

\bibitem{Criado:2018sdb}
J.~C. Criado and M.~P\'erez-Victoria, {\it {Field redefinitions in effective
  theories at higher orders}},  {\em JHEP} {\bf 03} (2019) 038,
  [\href{http://arxiv.org/abs/1811.09413}{{\tt arXiv:1811.09413}}].

\bibitem{Criado:2024mpx}
J.~C. Criado, J.~Jaeckel, and M.~Spannowsky, {\it {Field Redefinitions in
  Classical Field Theory with some Quantum Perspectives}},
  \href{http://arxiv.org/abs/2408.03369}{{\tt arXiv:2408.03369}}.

\bibitem{Coleman:1969sm}
S.~R. Coleman, J.~Wess, and B.~Zumino, {\it {Structure of phenomenological
  Lagrangians. 1.}},  {\em Phys. Rev.} {\bf 177} (1969) 2239--2247.

\bibitem{Callan:1969sn}
C.~G. Callan, Jr., S.~R. Coleman, J.~Wess, and B.~Zumino, {\it {Structure of
  phenomenological Lagrangians. 2.}},  {\em Phys. Rev.} {\bf 177} (1969)
  2247--2250.

\bibitem{Honerkamp:1971sh}
J.~Honerkamp, {\it {Chiral multiloops}},  {\em Nucl. Phys. B} {\bf 36} (1972)
  130--140.

\bibitem{Volkov:1973vd}
D.~V. Volkov, {\it {Phenomenological Lagrangians}},  {\em Fiz. Elem. Chast.
  Atom. Yadra} {\bf 4} (1973) 3--41.

\bibitem{Tataru:1975ys}
L.~Tataru, {\it {One Loop Divergences of the Nonlinear Chiral Theory}},  {\em
  Phys. Rev. D} {\bf 12} (1975) 3351--3352.

\bibitem{Alvarez-Gaume:1981exa}
L.~Alvarez-Gaume, D.~Z. Freedman, and S.~Mukhi, {\it {The Background Field
  Method and the Ultraviolet Structure of the Supersymmetric Nonlinear Sigma
  Model}},  {\em Annals Phys.} {\bf 134} (1981) 85.

\bibitem{Alvarez-Gaume:1981exv}
L.~Alvarez-Gaume and D.~Z. Freedman, {\it {Geometrical Structure and
  Ultraviolet Finiteness in the Supersymmetric Sigma Model}},  {\em Commun.
  Math. Phys.} {\bf 80} (1981) 443.

\bibitem{Vilkovisky:1984st}
G.~A. Vilkovisky, {\it {The Unique Effective Action in Quantum Field Theory}},
  {\em Nucl. Phys. B} {\bf 234} (1984) 125--137.

\bibitem{DeWitt:1984sjp}
B.~S. DeWitt, {\it {The spacetime approach to quantum field theory}},  in {\em
  {Les Houches Summer School on Theoretical Physics: Relativity, Groups and
  Topology}}, pp.~381--738, 1984.

\bibitem{Gaillard:1985uh}
M.~K. Gaillard, {\it {The Effective One Loop Lagrangian With Derivative
  Couplings}},  {\em Nucl. Phys. B} {\bf 268} (1986) 669--692.

\bibitem{DeWitt:1985sg}
B.~S. DeWitt, {\it {The Effective Action}},  in {\em {Les Houches School of
  Theoretical Physics: Architecture of Fundamental Interactions at Short
  Distances}}, pp.~1023--1058, 1987.

\bibitem{Georgi:1991ch}
H.~Georgi, {\it {On-shell effective field theory}},  {\em Nucl. Phys. B} {\bf
  361} (1991) 339--350.

\bibitem{Alonso:2015fsp}
R.~Alonso, E.~E. Jenkins, and A.~V. Manohar, {\it {A Geometric Formulation of
  Higgs Effective Field Theory: Measuring the Curvature of Scalar Field
  Space}},  {\em Phys. Lett. B} {\bf 754} (2016) 335--342,
  [\href{http://arxiv.org/abs/1511.00724}{{\tt arXiv:1511.00724}}].

\bibitem{Alonso:2016oah}
R.~Alonso, E.~E. Jenkins, and A.~V. Manohar, {\it {Geometry of the Scalar
  Sector}},  {\em JHEP} {\bf 08} (2016) 101,
  [\href{http://arxiv.org/abs/1605.03602}{{\tt arXiv:1605.03602}}].

\bibitem{Nagai:2019tgi}
R.~Nagai, M.~Tanabashi, K.~Tsumura, and Y.~Uchida, {\it {Symmetry and geometry
  in a generalized Higgs effective field theory: Finiteness of oblique
  corrections versus perturbative unitarity}},  {\em Phys. Rev. D} {\bf 100}
  (2019), no.~7 075020, [\href{http://arxiv.org/abs/1904.07618}{{\tt
  arXiv:1904.07618}}].

\bibitem{Cohen:2020xca}
T.~Cohen, N.~Craig, X.~Lu, and D.~Sutherland, {\it {Is SMEFT Enough?}},  {\em
  JHEP} {\bf 03} (2021) 237, [\href{http://arxiv.org/abs/2008.08597}{{\tt
  arXiv:2008.08597}}].

\bibitem{Cohen:2021ucp}
T.~Cohen, N.~Craig, X.~Lu, and D.~Sutherland, {\it {Unitarity violation and the
  geometry of Higgs EFTs}},  {\em JHEP} {\bf 12} (2021) 003,
  [\href{http://arxiv.org/abs/2108.03240}{{\tt arXiv:2108.03240}}].

\bibitem{Alonso:2021rac}
R.~Alonso and M.~West, {\it {Roads to the Standard Model}},  {\em Phys. Rev. D}
  {\bf 105} (2022), no.~9 096028, [\href{http://arxiv.org/abs/2109.13290}{{\tt
  arXiv:2109.13290}}].

\bibitem{Cheung:2021yog}
C.~Cheung, A.~Helset, and J.~Parra-Martinez, {\it {Geometric soft theorems}},
  {\em JHEP} {\bf 04} (2022) 011, [\href{http://arxiv.org/abs/2111.03045}{{\tt
  arXiv:2111.03045}}].

\bibitem{Helset:2022tlf}
A.~Helset, E.~E. Jenkins, and A.~V. Manohar, {\it {Geometry in scattering
  amplitudes}},  {\em Phys. Rev. D} {\bf 106} (2022), no.~11 116018,
  [\href{http://arxiv.org/abs/2210.08000}{{\tt arXiv:2210.08000}}].

\bibitem{Alonso:2023upf}
R.~Alonso, {\it {A primer on Higgs Effective Field Theory with Geometry}},
  \href{http://arxiv.org/abs/2307.14301}{{\tt arXiv:2307.14301}}.

\bibitem{Cohen:2022uuw}
T.~Cohen, N.~Craig, X.~Lu, and D.~Sutherland, {\it {On-Shell Covariance of
  Quantum Field Theory Amplitudes}},  {\em Phys. Rev. Lett.} {\bf 130} (2023),
  no.~4 041603, [\href{http://arxiv.org/abs/2202.06965}{{\tt
  arXiv:2202.06965}}].

\bibitem{Cheung:2022vnd}
C.~Cheung, A.~Helset, and J.~Parra-Martinez, {\it {Geometry-kinematics
  duality}},  {\em Phys. Rev. D} {\bf 106} (2022), no.~4 045016,
  [\href{http://arxiv.org/abs/2202.06972}{{\tt arXiv:2202.06972}}].

\bibitem{Cohen:2023ekv}
T.~Cohen, X.~Lu, and D.~Sutherland, {\it {On amplitudes and field
  redefinitions}},  {\em JHEP} {\bf 06} (2024) 149,
  [\href{http://arxiv.org/abs/2312.06748}{{\tt arXiv:2312.06748}}].

\bibitem{Craig:2023wni}
N.~Craig, Y.-T. Lee, X.~Lu, and D.~Sutherland, {\it {Effective field theories
  as Lagrange spaces}},  {\em JHEP} {\bf 11} (2023) 069,
  [\href{http://arxiv.org/abs/2305.09722}{{\tt arXiv:2305.09722}}].

\bibitem{Craig:2023hhp}
N.~Craig and Y.-T. Lee, {\it {Effective Field Theories on the Jet Bundle}},
  {\em Phys. Rev. Lett.} {\bf 132} (2024), no.~6 061602,
  [\href{http://arxiv.org/abs/2307.15742}{{\tt arXiv:2307.15742}}].

\bibitem{Alminawi:2023qtf}
M.~Alminawi, I.~Brivio, and J.~Davighi, {\it {Jet Bundle Geometry of Scalar
  Field Theories}},  {\em J. Phys. A} {\bf 57} (2024) 435401,
  [\href{http://arxiv.org/abs/2308.00017}{{\tt arXiv:2308.00017}}].

\bibitem{Cohen:2024bml}
T.~Cohen, X.~Lu, and Z.~Zhang, {\it {What is the Geometry of Effective Field
  Theories?}},  \href{http://arxiv.org/abs/2410.21378}{{\tt arXiv:2410.21378}}.

\bibitem{Lee:2024xqa}
Y.-T. Lee, {\it {Field Space Geometry and Nonlinear Supersymmetry}},
  \href{http://arxiv.org/abs/2410.21395}{{\tt arXiv:2410.21395}}.

\bibitem{Banta:2021dek}
I.~Banta, T.~Cohen, N.~Craig, X.~Lu, and D.~Sutherland, {\it {Non-decoupling
  new particles}},  {\em JHEP} {\bf 02} (2022) 029,
  [\href{http://arxiv.org/abs/2110.02967}{{\tt arXiv:2110.02967}}].

\bibitem{Helset:2020yio}
A.~Helset, A.~Martin, and M.~Trott, {\it {The Geometric Standard Model
  Effective Field Theory}},  {\em JHEP} {\bf 03} (2020) 163,
  [\href{http://arxiv.org/abs/2001.01453}{{\tt arXiv:2001.01453}}].

\bibitem{Hays:2020scx}
C.~Hays, A.~Helset, A.~Martin, and M.~Trott, {\it {Exact SMEFT formulation and
  expansion to $\mathcal{O}(v^4/\Lambda^4)$}},  {\em JHEP} {\bf 11} (2020) 087,
  [\href{http://arxiv.org/abs/2007.00565}{{\tt arXiv:2007.00565}}].

\bibitem{Derda:2024jvo}
M.~Derda, A.~Helset, and J.~Parra-Martinez, {\it {Soft scalars in effective
  field theory}},  {\em JHEP} {\bf 06} (2024) 133,
  [\href{http://arxiv.org/abs/2403.12142}{{\tt arXiv:2403.12142}}].

\bibitem{Helset:2022pde}
A.~Helset, E.~E. Jenkins, and A.~V. Manohar, {\it {Renormalization of the
  Standard Model Effective Field Theory from geometry}},  {\em JHEP} {\bf 02}
  (2023) 063, [\href{http://arxiv.org/abs/2212.03253}{{\tt arXiv:2212.03253}}].

\bibitem{Assi:2023zid}
B.~Assi, A.~Helset, A.~V. Manohar, J.~Pag\`es, and C.-H. Shen, {\it {Fermion
  Geometry and the Renormalization of the Standard Model Effective Field
  Theory}},  \href{http://arxiv.org/abs/2307.03187}{{\tt arXiv:2307.03187}}.

\bibitem{Jenkins:2023rtg}
E.~E. Jenkins, A.~V. Manohar, L.~Naterop, and J.~Pag\`es, {\it {An Algebraic
  Formula for Two Loop Renormalization of Scalar Quantum Field Theory}},
  \href{http://arxiv.org/abs/2308.06315}{{\tt arXiv:2308.06315}}.

\bibitem{Jenkins:2023bls}
E.~E. Jenkins, A.~V. Manohar, L.~Naterop, and J.~Pag\`es, {\it {Two Loop
  Renormalization of Scalar Theories using a Geometric Approach}},
  \href{http://arxiv.org/abs/2310.19883}{{\tt arXiv:2310.19883}}.

\bibitem{Henning:2014gca}
B.~Henning, X.~Lu, and H.~Murayama, {\it {What do precision Higgs measurements
  buy us?}},  \href{http://arxiv.org/abs/1404.1058}{{\tt arXiv:1404.1058}}.

\bibitem{Henning:2014wua}
B.~Henning, X.~Lu, and H.~Murayama, {\it {How to use the Standard Model
  effective field theory}},  {\em JHEP} {\bf 01} (2016) 023,
  [\href{http://arxiv.org/abs/1412.1837}{{\tt arXiv:1412.1837}}].

\bibitem{Henning:2016lyp}
B.~Henning, X.~Lu, and H.~Murayama, {\it {One-loop Matching and Running with
  Covariant Derivative Expansion}},  {\em JHEP} {\bf 01} (2018) 123,
  [\href{http://arxiv.org/abs/1604.01019}{{\tt arXiv:1604.01019}}].

\bibitem{Fuentes-Martin:2016uol}
J.~Fuentes-Martin, J.~Portoles, and P.~Ruiz-Femenia, {\it {Integrating out
  heavy particles with functional methods: a simplified framework}},  {\em
  JHEP} {\bf 09} (2016) 156, [\href{http://arxiv.org/abs/1607.02142}{{\tt
  arXiv:1607.02142}}].

\bibitem{Zhang:2016pja}
Z.~Zhang, {\it {Covariant diagrams for one-loop matching}},  {\em JHEP} {\bf
  05} (2017) 152, [\href{http://arxiv.org/abs/1610.00710}{{\tt
  arXiv:1610.00710}}].

\bibitem{Cohen:2019btp}
T.~Cohen, M.~Freytsis, and X.~Lu, {\it {Functional Methods for Heavy Quark
  Effective Theory}},  {\em JHEP} {\bf 06} (2020) 164,
  [\href{http://arxiv.org/abs/1912.08814}{{\tt arXiv:1912.08814}}].

\bibitem{Cohen:2020fcu}
T.~Cohen, X.~Lu, and Z.~Zhang, {\it {Functional Prescription for EFT
  Matching}},  {\em JHEP} {\bf 02} (2021) 228,
  [\href{http://arxiv.org/abs/2011.02484}{{\tt arXiv:2011.02484}}].

\bibitem{Cohen:2020qvb}
T.~Cohen, X.~Lu, and Z.~Zhang, {\it {STrEAMlining EFT Matching}},  {\em SciPost
  Phys.} {\bf 10} (2021), no.~5 098,
  [\href{http://arxiv.org/abs/2012.07851}{{\tt arXiv:2012.07851}}].

\bibitem{Fuentes-Martin:2020udw}
J.~Fuentes-Martin, M.~K\"onig, J.~Pag\`es, A.~E. Thomsen, and F.~Wilsch, {\it
  {SuperTracer: A Calculator of Functional Supertraces for One-Loop EFT
  Matching}},  {\em JHEP} {\bf 04} (2021) 281,
  [\href{http://arxiv.org/abs/2012.08506}{{\tt arXiv:2012.08506}}].

\bibitem{Fuentes-Martin:2022jrf}
J.~Fuentes-Mart\'\i{}n, M.~K\"onig, J.~Pag\`es, A.~E. Thomsen, and F.~Wilsch,
  {\it {A proof of concept for matchete: an automated tool for matching
  effective theories}},  {\em Eur. Phys. J. C} {\bf 83} (2023), no.~7 662,
  [\href{http://arxiv.org/abs/2212.04510}{{\tt arXiv:2212.04510}}].

\bibitem{Fuentes-Martin:2023ljp}
J.~Fuentes-Mart\'\i{}n, A.~Palavri\'c, and A.~E. Thomsen, {\it {Functional
  matching and renormalization group equations at two-loop order}},  {\em Phys.
  Lett. B} {\bf 851} (2024) 138557,
  [\href{http://arxiv.org/abs/2311.13630}{{\tt arXiv:2311.13630}}].

\bibitem{Born:2024mgz}
L.~Born, J.~Fuentes-Mart\'\i{}n, S.~Kvedarait\.{e}, and A.~E. Thomsen, {\it
  {Two-Loop Running in the Bosonic SMEFT Using Functional Methods}},
  \href{http://arxiv.org/abs/2410.07320}{{\tt arXiv:2410.07320}}.

\bibitem{Chan:1986jq}
L.-H. Chan, {\it {Derivative Expansion for the One Loop Effective Actions With
  Internal Symmetry}},  {\em Phys. Rev. Lett.} {\bf 57} (1986) 1199.

\bibitem{Cheyette:1987qz}
O.~Cheyette, {\it {Effective Action for the Standard Model With Large Higgs
  Mass}},  {\em Nucl. Phys. B} {\bf 297} (1988) 183--204.

\bibitem{Drozd:2015rsp}
A.~Drozd, J.~Ellis, J.~Quevillon, and T.~You, {\it {The Universal One-Loop
  Effective Action}},  {\em JHEP} {\bf 03} (2016) 180,
  [\href{http://arxiv.org/abs/1512.03003}{{\tt arXiv:1512.03003}}].

\bibitem{Ellis:2016enq}
S.~A.~R. Ellis, J.~Quevillon, T.~You, and Z.~Zhang, {\it {Mixed
  heavy\textendash{}light matching in the Universal One-Loop Effective
  Action}},  {\em Phys. Lett. B} {\bf 762} (2016) 166--176,
  [\href{http://arxiv.org/abs/1604.02445}{{\tt arXiv:1604.02445}}].

\bibitem{Ellis:2017jns}
S.~A.~R. Ellis, J.~Quevillon, T.~You, and Z.~Zhang, {\it {Extending the
  Universal One-Loop Effective Action: Heavy-Light Coefficients}},  {\em JHEP}
  {\bf 08} (2017) 054, [\href{http://arxiv.org/abs/1706.07765}{{\tt
  arXiv:1706.07765}}].

\bibitem{Summ:2018oko}
B.~Summ and A.~Voigt, {\it {Extending the Universal One-Loop Effective Action
  by Regularization Scheme Translating Operators}},  {\em JHEP} {\bf 08} (2018)
  026, [\href{http://arxiv.org/abs/1806.05171}{{\tt arXiv:1806.05171}}].

\bibitem{Kramer:2019fwz}
M.~Kr\"amer, B.~Summ, and A.~Voigt, {\it {Completing the scalar and fermionic
  Universal One-Loop Effective Action}},  {\em JHEP} {\bf 01} (2020) 079,
  [\href{http://arxiv.org/abs/1908.04798}{{\tt arXiv:1908.04798}}].

\bibitem{Ellis:2020ivx}
S.~A.~R. Ellis, J.~Quevillon, P.~N.~H. Vuong, T.~You, and Z.~Zhang, {\it {The
  Fermionic Universal One-Loop Effective Action}},  {\em JHEP} {\bf 11} (2020)
  078, [\href{http://arxiv.org/abs/2006.16260}{{\tt arXiv:2006.16260}}].

\bibitem{Angelescu:2020yzf}
A.~Angelescu and P.~Huang, {\it {Integrating Out New Fermions at One Loop}},
  {\em JHEP} {\bf 01} (2021) 049, [\href{http://arxiv.org/abs/2006.16532}{{\tt
  arXiv:2006.16532}}].

\bibitem{Larue:2023uyv}
R.~Larue and J.~Quevillon, {\it {The universal one-loop effective action with
  gravity}},  {\em JHEP} {\bf 11} (2023) 045,
  [\href{http://arxiv.org/abs/2303.10203}{{\tt arXiv:2303.10203}}].

\bibitem{Jenkins:2017jig}
E.~E. Jenkins, A.~V. Manohar, and P.~Stoffer, {\it {Low-Energy Effective Field
  Theory below the Electroweak Scale: Operators and Matching}},  {\em JHEP}
  {\bf 03} (2018) 016, [\href{http://arxiv.org/abs/1709.04486}{{\tt
  arXiv:1709.04486}}]. [Erratum: JHEP 12, 043 (2023)].

\bibitem{Craig:2021ksw}
N.~Craig, I.~G. Garcia, A.~Vainshtein, and Z.~Zhang, {\it {Magic zeroes and
  hidden symmetries}},  {\em JHEP} {\bf 05} (2022) 079,
  [\href{http://arxiv.org/abs/2112.05770}{{\tt arXiv:2112.05770}}].

\bibitem{Arkani-Hamed:2021xlp}
N.~Arkani-Hamed and K.~Harigaya, {\it {Naturalness and the muon magnetic
  moment}},  {\em JHEP} {\bf 09} (2021) 025,
  [\href{http://arxiv.org/abs/2106.01373}{{\tt arXiv:2106.01373}}].

\bibitem{DelleRose:2022ygn}
L.~Delle~Rose, B.~von Harling, and A.~Pomarol, {\it {Wilson coefficients and
  natural zeros from the on-shell viewpoint}},  {\em JHEP} {\bf 05} (2022) 120,
  [\href{http://arxiv.org/abs/2201.10572}{{\tt arXiv:2201.10572}}].

\bibitem{Bao:2024zzc}
Y.~Bao, J.~Gu, Z.~Liu, C.~Shu, and L.-T. Wang, {\it {Accidental Suppression of
  Wilson Coefficients in Higgs Coupling}},
  \href{http://arxiv.org/abs/2408.08948}{{\tt arXiv:2408.08948}}.

\bibitem{Alonso:2022ffe}
R.~Alonso and M.~West, {\it {On the effective action for scalars in a general
  manifold to any loop order}},  {\em Phys. Lett. B} {\bf 841} (2023) 137937,
  [\href{http://arxiv.org/abs/2207.02050}{{\tt arXiv:2207.02050}}].

\bibitem{Carmona:2021xtq}
A.~Carmona, A.~Lazopoulos, P.~Olgoso, and J.~Santiago, {\it {Matchmakereft:
  automated tree-level and one-loop matching}},  {\em SciPost Phys.} {\bf 12}
  (2022), no.~6 198, [\href{http://arxiv.org/abs/2112.10787}{{\tt
  arXiv:2112.10787}}].

\bibitem{Georgi:1993mps}
H.~Georgi, {\it {Effective field theory}},  {\em Ann. Rev. Nucl. Part. Sci.}
  {\bf 43} (1993) 209--252.

\bibitem{Beneke:1997zp}
M.~Beneke and V.~A. Smirnov, {\it {Asymptotic expansion of Feynman integrals
  near threshold}},  {\em Nucl. Phys. B} {\bf 522} (1998) 321--344,
  [\href{http://arxiv.org/abs/hep-ph/9711391}{{\tt hep-ph/9711391}}].

\bibitem{Smirnov:2002pj}
V.~A. Smirnov, {\it {Applied asymptotic expansions in momenta and masses}},
  {\em Springer Tracts Mod. Phys.} {\bf 177} (2002) 1--262.

\bibitem{Haisch:2020ahr}
U.~Haisch, M.~Ruhdorfer, E.~Salvioni, E.~Venturini, and A.~Weiler, {\it
  {Singlet night in Feynman-ville: one-loop matching of a real scalar}},  {\em
  JHEP} {\bf 04} (2020) 164, [\href{http://arxiv.org/abs/2003.05936}{{\tt
  arXiv:2003.05936}}]. [Erratum: JHEP 07, 066 (2020)].

\bibitem{Jiang:2018pbd}
M.~Jiang, N.~Craig, Y.-Y. Li, and D.~Sutherland, {\it {Complete one-loop
  matching for a singlet scalar in the Standard Model EFT}},  {\em JHEP} {\bf
  02} (2019) 031, [\href{http://arxiv.org/abs/1811.08878}{{\tt
  arXiv:1811.08878}}]. [Erratum: JHEP 01, 135 (2021)].

\bibitem{deBlas:2017xtg}
J.~de~Blas, J.~C. Criado, M.~Perez-Victoria, and J.~Santiago, {\it {Effective
  description of general extensions of the Standard Model: the complete
  tree-level dictionary}},  {\em JHEP} {\bf 03} (2018) 109,
  [\href{http://arxiv.org/abs/1711.10391}{{\tt arXiv:1711.10391}}].

\bibitem{Li:2023cwy}
X.-X. Li, Z.~Ren, and J.-H. Yub, {\it {Complete tree-level dictionary between
  simplified BSM models and SMEFT d\ensuremath{\leq}7 operators}},  {\em Phys.
  Rev. D} {\bf 109} (2024), no.~9 095041,
  [\href{http://arxiv.org/abs/2307.10380}{{\tt arXiv:2307.10380}}].

\bibitem{Alonso:2014rga}
R.~Alonso, E.~E. Jenkins, and A.~V. Manohar, {\it {Holomorphy without
  Supersymmetry in the Standard Model Effective Field Theory}},  {\em Phys.
  Lett. B} {\bf 739} (2014) 95--98, [\href{http://arxiv.org/abs/1409.0868}{{\tt
  arXiv:1409.0868}}].

\bibitem{Elias-Miro:2014eia}
J.~Elias-Miro, J.~R. Espinosa, and A.~Pomarol, {\it {One-loop
  non-renormalization results in EFTs}},  {\em Phys. Lett. B} {\bf 747} (2015)
  272--280, [\href{http://arxiv.org/abs/1412.7151}{{\tt arXiv:1412.7151}}].

\bibitem{Cheung:2015aba}
C.~Cheung and C.-H. Shen, {\it {Nonrenormalization Theorems without
  Supersymmetry}},  {\em Phys. Rev. Lett.} {\bf 115} (2015), no.~7 071601,
  [\href{http://arxiv.org/abs/1505.01844}{{\tt arXiv:1505.01844}}].

\bibitem{Bern:2019wie}
Z.~Bern, J.~Parra-Martinez, and E.~Sawyer, {\it {Nonrenormalization and
  Operator Mixing via On-Shell Methods}},  {\em Phys. Rev. Lett.} {\bf 124}
  (2020), no.~5 051601, [\href{http://arxiv.org/abs/1910.05831}{{\tt
  arXiv:1910.05831}}].

\bibitem{Craig:2019wmo}
N.~Craig, M.~Jiang, Y.-Y. Li, and D.~Sutherland, {\it {Loops and Trees in
  Generic EFTs}},  {\em JHEP} {\bf 08} (2020) 086,
  [\href{http://arxiv.org/abs/2001.00017}{{\tt arXiv:2001.00017}}].

\bibitem{Jiang:2020rwz}
M.~Jiang, J.~Shu, M.-L. Xiao, and Y.-H. Zheng, {\it {Partial Wave Amplitude
  Basis and Selection Rules in Effective Field Theories}},  {\em Phys. Rev.
  Lett.} {\bf 126} (2021), no.~1 011601,
  [\href{http://arxiv.org/abs/2001.04481}{{\tt arXiv:2001.04481}}].

\bibitem{EliasMiro:2020tdv}
J.~Elias~Mir\'o, J.~Ingoldby, and M.~Riembau, {\it {EFT anomalous dimensions
  from the S-matrix}},  {\em JHEP} {\bf 09} (2020) 163,
  [\href{http://arxiv.org/abs/2005.06983}{{\tt arXiv:2005.06983}}].

\bibitem{Baratella:2020lzz}
P.~Baratella, C.~Fernandez, and A.~Pomarol, {\it {Renormalization of
  Higher-Dimensional Operators from On-shell Amplitudes}},  {\em Nucl. Phys. B}
  {\bf 959} (2020) 115155, [\href{http://arxiv.org/abs/2005.07129}{{\tt
  arXiv:2005.07129}}].

\bibitem{Bern:2020ikv}
Z.~Bern, J.~Parra-Martinez, and E.~Sawyer, {\it {Structure of two-loop SMEFT
  anomalous dimensions via on-shell methods}},  {\em JHEP} {\bf 10} (2020) 211,
  [\href{http://arxiv.org/abs/2005.12917}{{\tt arXiv:2005.12917}}].

\bibitem{Cao:2021cdt}
W.~Cao, F.~Herzog, T.~Melia, and J.~R. Nepveu, {\it {Renormalization and
  non-renormalization of scalar EFTs at higher orders}},  {\em JHEP} {\bf 09}
  (2021) 014, [\href{http://arxiv.org/abs/2105.12742}{{\tt arXiv:2105.12742}}].

\bibitem{Chala:2023jyx}
M.~Chala, {\it {Constraints on anomalous dimensions from the positivity of the
  S matrix}},  {\em Phys. Rev. D} {\bf 108} (2023), no.~1 015031,
  [\href{http://arxiv.org/abs/2301.09995}{{\tt arXiv:2301.09995}}].

\bibitem{Cao:2023adc}
W.~Cao, F.~Herzog, T.~Melia, and J.~Roosmale~Nepveu, {\it {Non-linear
  non-renormalization theorems}},  {\em JHEP} {\bf 08} (2023) 080,
  [\href{http://arxiv.org/abs/2303.07391}{{\tt arXiv:2303.07391}}].

\bibitem{Chala:2023xjy}
M.~Chala and X.~Li, {\it {Positivity restrictions on the mixing of
  dimension-eight SMEFT operators}},  {\em Phys. Rev. D} {\bf 109} (2024),
  no.~6 065015, [\href{http://arxiv.org/abs/2309.16611}{{\tt
  arXiv:2309.16611}}].

\bibitem{Finn:2019aip}
K.~Finn, S.~Karamitsos, and A.~Pilaftsis, {\it {Frame Covariance in Quantum
  Gravity}},  {\em Phys. Rev. D} {\bf 102} (2020), no.~4 045014,
  [\href{http://arxiv.org/abs/1910.06661}{{\tt arXiv:1910.06661}}].

\bibitem{Finn:2020nvn}
K.~Finn, S.~Karamitsos, and A.~Pilaftsis, {\it {Frame covariant formalism for
  fermionic theories}},  {\em Eur. Phys. J. C} {\bf 81} (2021), no.~7 572,
  [\href{http://arxiv.org/abs/2006.05831}{{\tt arXiv:2006.05831}}].

\bibitem{Pilaftsis:2022las}
A.~Pilaftsis, K.~Finn, V.~Gattus, and S.~Karamitsos, {\it {Geometrising the
  Micro-Cosmos on a Supermanifold}},  {\em PoS} {\bf CORFU2021} (2022) 080,
  [\href{http://arxiv.org/abs/2204.00123}{{\tt arXiv:2204.00123}}].

\bibitem{Gattus:2023gep}
V.~Gattus and A.~Pilaftsis, {\it {Minimal supergeometric quantum field
  theories}},  {\em Phys. Lett. B} {\bf 846} (2023) 138234,
  [\href{http://arxiv.org/abs/2307.01126}{{\tt arXiv:2307.01126}}].

\end{thebibliography}\endgroup

\end{document}